\documentclass[preprint,showpacs,preprintnumbers,amsmath,amssymb]{revtex4-2}

\usepackage{graphicx}
\usepackage{dcolumn}

\usepackage{float}
\usepackage{color}
\usepackage{xcolor}
\usepackage[T1]{fontenc}       

\usepackage{notes2bib}
\usepackage{caption}
\usepackage{subcaption}
\usepackage{ragged2e}
\usepackage{booktabs}
\usepackage{ulem}

\begin{document}


\title{Calculated iron $L_{2,3}$ x-ray absorption and XMCD of spin-crossover Fe(phen)$_{2}$(NCS)$_{2}$ molecule adsorbed on Cu(001) surface}
\date{\today}

\author{R. Pasquier and M. Alouani}
\affiliation{%
Universit\'{e} de Strasbourg, 
Institut de Physique et de Chimie des Mat\'{e}riaux de Strasbourg, CNRS-UNISTRA UMR 7504, 67034 Strasbourg, France}%

\begin{abstract}
    The projector augmented wave method 
    has been used to compute the iron 
    L$_{2,3}$ edges of x-ray absorption spectra (XAS) and x-ray magnetic circular
    dichroism (XMCD) of the spin-crossover Fe(phen)$_{2}$(NCS)$_{2}$ molecule  when adsorbed
    on Cu(001) surface and in the gas phase, for both the high spin (HS) and low spin (LS) states. 
    The electronic structures of both HS and LS states have been  calculated using 
    the spin-polarized generalized gradient approximation  for the exchange-correlation
    potential, and  the strongly 
    localized iron $3d$ states are described using Dudarev's rotationally invariant 
    formulation of the DFT+U method.  
    It is shown that only the iron  all-electron partial
    waves are necessary to calculate the  XAS transition matrix elements
    in the electric dipole approximation, as the contribution of the pseudo
    partial waves is compensated by the plane wave component of the wave function. 
    {It is  found that the calculated XAS and XMCD  with the static core hole or the Slater 
    transition state half hole are in less good
    agreement with experiment than those using the so called initial state. This disagreement is due to  
    the reduction of the iron spin magnetic moment caused by the static screening of the core hole by  the 
    photo-electron. The L$_{2,3}$ XAS formula is found to be directly related to the unoccupied $3d$ density of 
    states (DOS), and hence 
    the symmetry broken  $e_g$ and the $t_{2g}$  iron DOS}   are used to explain the XAS and XMCD results.
    It is  demonstrated  that the dependence of the HS  XMCD on the direction of incident 
    x-ray circularly polarized light with respect to the magnetization direction can be used to determine 
    the iron octahedron  deformation, while the  XMCD for various  magnetization directions 
    is directly related to the anisotropy of the orbital magnetic moment and the magneto-crystalline energy.
    The Thole-Carra-Van der Laan XMCD sum rules have been applied to the  XMCD L$_{2,3}$ spectra 
    to compute the spin and orbital magnetic moments.  It is
    shown that the magnetic dipole moment $T_z$  is very large  due to  the
    strong distortion of the iron octahedron and is  necessary  for an accurate
    determination of the sum rule computed spin magnetic moment. 
\end{abstract}

\maketitle
\newpage

\section{Introduction}


Mastering the physics and chemistry of  isolated bistable molecules that are adsorbed 
on surfaces  is necessary to  further advance  
information technology as they  have potential applications 
for display devices,  data storage, and  organic transistors. 
 For example, 
the spin state of spin-crossover (SCO) molecules can be switched between the low-spin (LS) and
high-spin (HS) states 
by an external perturbation such as a variation in the  
temperature, light, pressure, magnetic or electric field \cite{Gultlich,Real,Hao,Bousseksou,Jeftic,Ksenofontov,Tarafder}.
This artificially triggered $d-d$ electronic transition originates from the redistribution of 
the transition metal 3$d$ electrons between the symmetry broken  $e_g$ and $t_{2g}$  orbitals 
due to the structural deformation  of the octahedral transition-metal complex. 

X-ray absorption spectroscopy (XAS) and x-ray magnetic circular dichroism (XMCD)  techniques
have been extensively used to study SCO molecular systems. 
The absorption spectra of the HS and LS states are quite distinct due to 
their different  magnetic ground state properties\cite{Rudd2005-mr,briois1995,Lee2000,Hocking2006,Miyamachi2012}. 
These techniques are therefore 
powerful characterization tools for studying  the spin state of SCO compounds in all magnetic phases. 
Naturally, such spectra have been computed for the prototype
Fe(phen)$_{2}$(NCS)$_{2}$ (Fephen) spin-crossover molecule, where phen is the 1,10-phenanthroline,   
and they are commonly used as a reference by experimentalists working on these systems \cite{briois1995}. 
Miyamachi \textit{et al.}, for example, studied  the spin-crossover phenomenon in
a Fephen system both in the gas phase and adsorbed on a Cu(001) surface along with its XAS and  
and XMCD \cite{Miyamachi2012}.
Among other things, a Fano resonance in the conductance at  zero-bias was observed only in the HS state. 
The origin of this anomalous conductance  has
been arduous to elucidate due to the subtle electronic coupling between the
surface and the complex, and it has led to an  extensive body of work \cite{gruber:hal-01883180}. 

From a fundamental perspective, the transition from the HS to LS in Fe(II) complexes
involves the spin transition 
$ (t_{2g})^{3\uparrow} (e_g)^{2\uparrow}  
(t_{2g})^{1\downarrow}  \rightarrow (t_{2g})^{3\uparrow} (t_{2g})^{3\downarrow}  (e_g)^{0} $ 
(see Ref. \cite{Pasquier2022}).
This spin transition   stabilizes  the low-spin state because of    the full
occupation of the low-lying symmetry broken $t_{2g} $ energy levels. It is therefore 
interesting to compute the LS and HS  XAS L$_{2,3}$ edges from first principles  and 
compare them directly  to experiment to validate this spin transition.
To this end,   we have calculated the x-ray absorption
spectra and XMCD of the Fephen molecule both on the gas phase and  adsorbed on a Cu(001) surface 
in the electric dipole approximation and have compared our findings  to experiment\cite{Miyamachi2012}. 
{We have shown, in particular, that the XAS and  XMCD calculations
including  the static core hole  are in a worse agreement with experiment 
than the calculations using the ground state. 
This fact remains true even when using the Slater transition rule, where only  a half hole is used in  
the $2p$ core states\cite{Slater1972,PhysRevB.5.844}. This disagreement is explained below in terms of the 
reduction of spin moment caused by the additional screening of the core hole by the photo-electron.}
In addition, we have  demonstrated  that the dependance of the  XMCD signal  on the direction of incident 
circularly polarized 
light  can be used to determine the deformation of the iron octahedron, and 
the XMCD for various magnetization orientations gives the
anisotropy of the orbital magnetic moment which is  related to the magneto-crystalline energy (MCA). 
Those findings were not investigated in 
Ref. \cite{Miyamachi2012} but are very useful for the characterization of SCO molecules adsorbed 
on a metallic substrate.

To understand the distinct characteristics  associated with absorption spectra of the 
HS and LS,  we have  first shown that the XAS is related to the $3d$ density of states 
and the XAS results can therefore  be explained in terms 
of the density of states of the $3d$-electrons. We have therefore 
compared the  symmetry broken parent $e_g$ and $t_{2g}$  density of states to the calculated
XAS results.\cite{egt2g}  We have  also tested the validity
of the so-called XMCD sum rules for molecular systems and shown  that it is  
necessary  to take into account the contribution of the  magnetic dipole operator $T_z$
to obtain spin magnetic moments in agreement with those obtained from the 
electronic structure calculation or experiment. Finally, we have derived the plane wave
contribution to the x-ray electronic transition matrix elements within the
projected augmented plane wave (PAW) method \cite{PhysRevB.50.17953} and have shown that it has a negligible effect on  the XAS and XMCD.

This paper is organized as follows: In the second section, we provide a brief description of  our 
method of calculation and implementation 
of the XAS and the XMCD in \textsc{VASP} and show that the dependence of the XMCD   on the direction of the 
incident circularly polarized 
x-ray beam is directly related to the distortion of the iron octahedron. In addition,  using the XMCD sum rules 
we show  that the  XMCD for different magnetization 
directions can be used
to determine the  anisotropy of the orbital magnetic moment.   {In a third section,  we 
present our results for the XAS and XMCD at the iron $L_{2,3}$ edges  for both the gas phase and 
the molecule adsorbed on a  Cu(001) surface and compare them to the available
experimental data and show that the calculation without static core hole is in better agreement
with experiment.}  We then give an interpretation of the HS and LS XAS and XMCD in terms of  
symmetry broken $e_g$  and  $t_{2g}$ density of states.
At the end of this  section,  we demonstrate how to utilize the XMCD sum rules to compute the spin and 
orbital magnetic moments as well as   the importance   
of the magnetic dipole moment for the determination of the  spin magnetic moment. 
The derivation of the plane wave contribution to the x-ray matrix elements, the implementation of  the 
magnetic dipole moment in the \textsc{VASP} package, as well as the approximation of the XAS by weighted 
partial density of states of the conduction electrons are provided in the appendices.

\section{Computational method}
\subsection{Computational details}

The electronic structure is computed using \textsc{VASP} (Vienna Ab initio
Simulation) package\cite{PhysRevB.47.558,PhysRevB.49.14251,KRESSE199615,PhysRevB.54.11169,PhysRevB.59.1758},
 which implements the Kohn-Sham density functional theory
(DFT) within the \textsc{PAW} method\cite{PhysRevB.50.17953}.
The spin-polarized generalized gradient approximation (GGA) with the
Perdew-Burke-Ernzerhof (PBE) functional \cite{PhysRevLett.77.3865} is employed to
describe the exchange-correlation potential, whereas the Van der Waals
interaction, which is relevant mostly between the surface and the molecule, is
taken into account using the semi-empirical Grimme approximation at the DFT-D2
level \cite{https://doi.org/10.1002/jcc.20495}. Note that we have attempted  to use
the more accurate DFT-D3 and DFT-D4 methods, but we have encountered convergence issues.
The strongly localized iron $3d$ states are described using Dudarev's
rotationally invariant formulation of Liechtenstein's DFT+U method
\cite{PhysRevB.52.R5467,PhysRevB.57.1505}. The value of the effective Hubbard
parameter is $U-J=2.1$ eV, which accurately reflects the energy difference
between the two spin states of Fe(II) \cite{PhysRevB.87.144413}. The total
energy is converged to $10^{-5}$ eV and the plane wave cutoff is set at $500$
eV. The ionic force relaxation threshold is set at $5.10^{-2}$ eV/\AA $~~$ in
each direction. We have used a  supercell with dimensions 20.4$\times$20.4$\times$30
\AA$^3$. The surface is simulated using a $8\times 8\times 3$ stack of copper
oriented along the (001) direction and  contains 192 atoms. The Fermi energy has been
calculated
using a Gaussian smearing with a width of $0.1$ eV.  This entropy has been removed
from the computed total energy. The calculations are carried at the $\Gamma$
point only given the large dimensions of the supercell. 

\subsection{X-ray absorption and XMCD}

One can define the x-ray absorption cross-section with polarization $\mu$ in a
general way using Fermi's golden rule \cite{doi:10.1098/rspa.1927.0039}:
\begin{equation}
\label{eq:XASGoldRule}
\sigma^{\mu}(\omega) = \frac{4\pi \alpha \hbar}{m_e^2 \omega} \sum_{if}\left|\langle f|p_{\mu}|i\rangle \right|^2 \delta(\hbar \omega-\epsilon_f + \epsilon_i ).
\end{equation}
Here, $\alpha$ is the fine-structure constant, $m_e$ is the electron mass, $i$
and $f$ stand respectively for the core states and the conduction states, and 
their energies $\epsilon_i$ and $\epsilon_f$). Here $p_{\mu}=-i\hbar
\nabla_{\mu}$ is the projection of the momentum operator on the
$\{\mu=-1,0,1\}$ polarization direction. These directions, along with the
corresponding cross-section, are defined as:
\begin{align}
\label{polar}
    \sigma^{\mu=\pm} &:  p_{\mu=\pm} = \frac{\mp 1}{\sqrt{2}}\left(p_x\pm i p_y\right) \\    
    \nonumber \sigma^{\mu=0}  &: p_{\mu=0} = p_z.
\end{align}

{There are several ways to calculate the x-ray absorption cross-section,
ranging from the analytical evaluation of transition matrix elements
\cite{PhysRevB.52.12766} to core-hole \cite{PhysRevB.98.235205} or ligand field
DFT methods \cite{https://doi.org/10.1002/qua.26081}. In this study, our
approach is based on  the \textsc{PAW} method  within the 
DFT calculations to compute these matrix elements, and it has been  already used to
compute the K and L$_{2,3}$ edges in iron to achieve quantitative agreement with experimental data
\cite{DIXIT2016136}. Here, we extend our method by including the plane wave
contribution and thus enabling the computation of the XMCD spectra for any direction of the
magnetization and any direction of the incident circularly polarized x-ray.  We have also 
determined the XAS in terms of the  partial density of states of the probed atom. This will be
used later to analyze the different features in the XAS in terms of the symmetry broken $e_g$ and
$t_{2g}$ density of states of the ideal octahedron. Note that in this formalism, its straightforward to include
the effect of a static core hole or to use the Slater transition rule\cite{Slater1972,PhysRevB.5.844}, 
where only one half core state
is included. However, the static core hole usually only slightly improves the K-edge spectrum, but it 
seldom leads to  improved  L$_{2,3}$ edges
compared to  calculations with the initial state\cite{Calandra2013}. Indeed, we  will show 
below that  the initial state calculation agrees  better with experiment than 
calculations using a static full core hole or Slater's half hole.} \\ 

Within \textsc{PAW} \cite{PhysRevB.50.17953}, the core states
are  considered frozen and kept unchanged in the pseudopotential files, and they
are usually fully relativistic $|J,M\rangle$, i.e.,  solutions to the Dirac 
equation \cite{doi:10.1098/rspa.1928.0023}.
This  means that we need to work in a $|JMLS\rangle $ coupled basis set:
\begin{equation}
\label{JMLS}
    |i\rangle =|J,M\rangle =\sum_{m',s} \langle \ell',m',1/2,s|J,M\rangle  |\ell',m',1/2,s\rangle =\sum_{m',s} C^{J,M}_{\ell',m',1/2,s} |\ell',m',1/2,s\rangle ,
\end{equation}
where $\ell$, $m$ and $s$  the usual angular momentum and spin quantum numbers, 
 the  $C^{J,M}_{\ell',m',1/2,s}$ are the usual Clebsch-Gordan coefficients and $\ell^\prime = 1$ for the $L_{2,3}$ edges. It
should be noted that we disregard the contribution of the minor part of the
Dirac bispinor when computing the matrix elements because the conduction states are scalar  
relativistic and the small component  contribution is negligible.  
The conduction states are the computed Kohn-Sham
orbitals $|n,\mathbf{k},s\rangle $, which can be written in the \textsc{PAW} method as:
\begin{equation}
\label{KSPAW}
    |f\rangle =|n,\mathbf{k},s\rangle =|\widetilde{n,\mathbf{k},s}\rangle +\sum_{p,\ell,m}\widetilde{P}^{n,\mathbf{k},s}_{p,\ell,m}(|p,\ell,m,s\rangle -|\widetilde{p,\ell,m,s}\rangle ),
\end{equation}
where $n$ is the band index, $\mathbf{k}$ the wavevector and $s$  the spin
index and $\widetilde{P}^{n,\mathbf{k},s}_{p,\ell,m}$ is the projection value of the 
pseudo Kohn-Sham wave functions  
 on the \textsc{PAW} projector functions (for more details see Ref. \cite{PhysRevB.50.17953}). 
Here $p$ is used for multiple projector functions to improve the atomic basis set. 
Usually $p$ is limited to one or two projector functions  per angular momentum $\ell$. \\

{It should  be noted that our implementation can include only the static  
core-hole effects using a supercell geometry, unlike other \textsc{PAW} implementations, such as the 
Taillefumier \textsl{et al.}
method \cite{PhysRevB.66.195107}  where a continued fraction formulation was used to compute the 
K-edge x-ray absorption near-edge structures in presence of 
a core hole. However, our calculations do not include  multiplet structures and dynamical core-hole screening. 
Although
this appears to be  a drastic approximation,  we will show that our implementation  
is sufficient to obtain qualitative agreement with experiment. The
$|\widetilde{n,\mathbf{k},s}\rangle $ are the so-called pseudo wave functions
associated with the pseudo-partial waves $|\widetilde{p,\ell,m,s}\rangle $,
whereas the $|p,\ell,m,s\rangle $ are the all electron partial waves. The
pseudo and plane wave contributions will be shown to be negligible as the
3$d$ electrons of iron are strongly localized within the augmentation region,
and these corrections are hence extremely small up to several dozens of eV
above the Fermi level.  They are  therefore only relevant for EXAFS, which is not
the subject of this work. This naturally leads 
 us to limit the calculation to the relevant photo-electron energy range 
when not including the plane wave
contribution. These partial waves are indexed by $p$ the projector index, and
$\ell$, $m$ and $s$  the usual angular momentum and spin quantum numbers, with
$P^{n,\mathbf{k},s}_{p,\ell,m}$ the associated projection value of the
Kohn-Sham pseudo wave functions:}
\begin{equation}
\label{KSparproj}
    P^{n,\mathbf{k},s}_{p,\ell,m}=\; \langle g_{p,\ell,m,s}|\widetilde{n,\mathbf{k},s}\rangle,
\end{equation}
where $g_{p,\ell,m,s}$ are the usual $\textsc{PAW}$ projector functions.
Note that \textsc{VASP} uses cubic harmonics $\mathcal{Y}_\ell^m$, whereas the formula is
computed for spherical harmonics ${Y}_\ell^m$. We should then transform 
the projections back into the spherical harmonics basis when doing the actual
computation by using the usual unitary transformation $U$ from cubic to spherical harmonics.
We can show that it amounts in writing the projections as 
$P^{n,\mathbf{k},s}_{p}=U^{-1}*\widetilde{P}^{n,\mathbf{k},s}_{p}$, 
where $\widetilde{P}^{n,\mathbf{k},s}_{p}$ represents the vector of the cubic projections, as given by Eq. 
\ref{KSPAW}, that are computed by \textsc{VASP}.
\\\\
Using these formulas together  with the golden rule, we can find:
\begin{equation}
\label{eq:secXASGoldRule1}
\sigma^{\mu}(\omega) = \frac{4\pi \alpha \hbar}{m_e^2 \omega} \sum_{M,n,\mathbf{k},s}
    \bigg|\sum_{p,\ell,m,m'}C^{J,M}_{\ell',m',1/2,s}
    \langle 
    p,\ell,m
    |p_{\mu}|
    \ell',m'
    \rangle  
    P^*{}^{n,\mathbf{k},s}_{p,\ell,m}\bigg|^2 
    \delta(\hbar \omega-\epsilon_{n \mathbf{k} s} + \epsilon_{JM} ),
\end{equation}
where we have used the fact that the spin is conserved by the momentum operator.
The $\epsilon_{n \mathbf{k} s}$ and $\epsilon_{JM}$ are respectively the
Kohn-Sham eigenvalues and the relativistic core energies. Note that \textsc{VASP} does
not compute the spin-orbit splitting between the $(J-1/2)$ and $(J+1/2)$ core
states, and we have therefore taken this splitting from the result of a
relativistic all electron atomic program calculation\cite{Koelling_1977}. This also implies that the spectra
$\sigma^{\mu}(\omega)$ are $J$ dependent, although this will be kept implicit
in our notations. \\ 
Using Wigner-Eckart's theorem
\cite{sakurai_napolitano_2020,Edmonds+2016}, one can then show that
\begin{equation}
\label{WigEck}
    \langle 
    p,\ell,m |p_{\mu}| \ell',m'
    \rangle =\frac{C^{\ell,m}_{\ell',m',1,\mu}}{C^{\ell,0}_{\ell',0,1,0}}
    \langle p,\ell,0 |p_{0}| \ell',0 \rangle .
\end{equation}
We therefore recover the so-called dipolar selection rules: $\ell=\ell'\pm 1$ and
$m=\mu+m'$.  Using angular momentum algebra \cite{Edmonds+2016}, we have the
following closed formula that will be used to compute the reduced matrix element
for each projector $p$: 
\begin{equation}
\label{integWigEck}
    \begin{aligned}
        \langle 
        p,\ell, 0
        \left|\nabla_{0}\right| 
        \ell^{\prime}, 0
        \rangle=& 
        \delta_{ \ell, \ell^{\prime}+1 } 
        \frac{\ell}{\sqrt{\left(2 \ell-1\right)\left(2 \ell+1\right)}}
        \left[\left(\phi_{p,\ell}\left|\partial_{r}\right| \phi^c_{\ell'}\right)-
        \left(\ell -1\right)\left(\phi_{p,\ell}\left|r^{-1}\right| \phi^c_{\ell'} \right)\right] \\
&+\delta_{\ell,\ell^{\prime}-1} 
\frac{\ell+1}{\sqrt{\left(2 \ell+1\right)\left(2 \ell+3\right)}}
\left[\left(\phi_{p,\ell}\left|\partial_{r}\right| \phi^c_{\ell'}\right)+\left(\ell+2\right)
        \left(\phi_{p,\ell}\left|r^{-1}\right| \phi^c_{\ell'}\right)\right],
\end{aligned}
\end{equation}
where we introduced the radial functions associated with the core $\phi^c$ or
the conduction states $\phi$, along with the radial integration
$\left(\phi_{p,\ell}\left|r^{\alpha}\right| \phi^c_{\ell'}\right)=\int
\mathrm{d}r  \phi_{p,\ell}(r) r^{\alpha+2}  \phi^c_{\ell'}(r)$. Note that for
most weakly relativistic systems such that $\alpha^2 Z^2 << 1$ (including iron,
where $\alpha^2 Z^2 \approx 0.03 $), the fine structure corrections to the
radial eigenfunctions are very small \cite{GravedePeralta2020} and consequently one
can safely use non-relativistic radial wave functions for the core states. 
However, \textsc{VASP} allows us to compute the
relativistic radial wave functions by solving Dirac's equation, so we will use them.  

\begin{figure}[H]
\centering{
\begin{tabular}{c}
\includegraphics[scale=0.3]{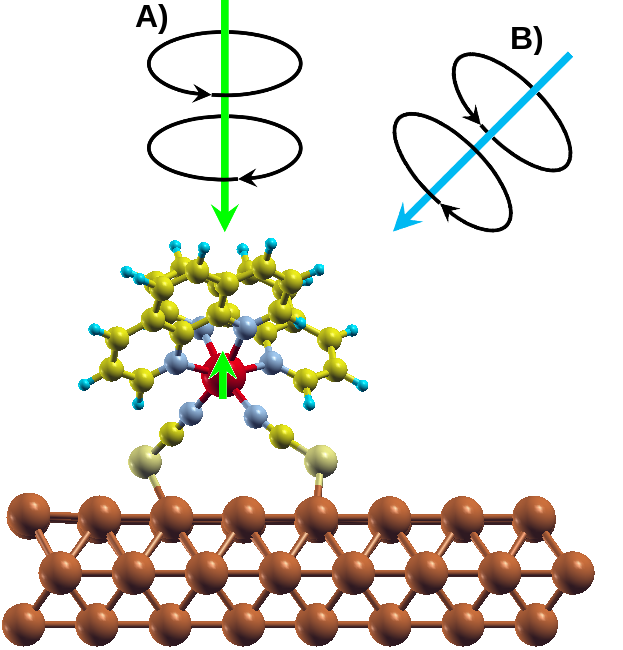}  \\
\end{tabular}
}

\caption{ Fephen  molecule 
    on top of the Cu(001) surface. 
    The direction of the incident circularly polarized light is along the $z$-direction (A) and along $\vartheta = \varphi = \pi/4$ (B).
    The spin magnetic moment direction  is set  along the $z$-direction, perpendicular to the Cu(001) surface,  
    when the spin-orbit coupling is included.}
\label{fig:polarization}
\end{figure}

To conclude, in the momentum representation, the following expression for the absorption spectrum can be shown:
\begin{equation}
\label{pXAS}
\sigma^{\mu}(\omega)=\frac{4\pi \alpha \hbar^3}{m_e^2 \omega} 
    \sum_{M,n,k,s} \left|\sum_{p,\ell,m,m'} 
    C^{J,M}_{\ell',m',1/2,s} 
    \frac{C^{\ell,m}_{\ell',m',1,\mu}}{C^{\ell,0}_{\ell',0,1,0}}
    \langle p,\ell,0\left|\nabla_0\right|\ell',0\rangle P^*{}^{n,\mathbf{k},s}_{p,\ell,m} \right|^2 
    \delta \left(\hbar \omega - \epsilon_{n \mathbf{k} s} + \epsilon_{JM} \right).
\end{equation}
The polarization is defined as in equation (\ref{polar}), and the XAS and XMCD
corresponding respectively to the 
$\sigma_{XAS}=\frac{1}{3}(\sigma^0+\sigma^{-}+\sigma^{+})$ and $\sigma_{XMCD}=\sigma^{+}-\sigma^{-}$ spectra 
are computed using Eq. \ref{pXAS}. 
    {The matrix elements $\langle p,\ell,0\left|\nabla_0\right|\ell',0\rangle $ are
    computed using equation (\ref{integWigEck}), and the radial integrations are
    cut at the augmentation radius for consistency. However, given the low
    symmetry of the molecule under study, it is important to note that the dependence of
    the XMCD signal   on the direction of the incident circularly polarized
    light is a signature of the distortion of the iron octahedron.  Here we have
    used a global coordinates system (O, $x,\; y,\; z$) and have assumed that
    the direction of incident light is given by the two spherical angles
    ($\vartheta$, $\varphi$). We can  therefore write the cross-section
    $\sigma_{XMCD}$ for any incident light direction specified by $\vartheta$, $\varphi$
    as shown in Fig. \ref{fig:polarization}. The figure  shows also  that the spin
    quantization direction is  fixed along a given direction, as it would
    be done experimentally with a magnetic field. Here  we take the (001) direction
    as a reference. It will be shown later that this direction corresponds to the
    lowest total energy when the spin-orbit coupling is included. We need
    therefore to rotate the matrix elements from the local frame of reference
    (O, $ x^\prime, y^\prime, z^\prime$), where the $z^\prime$ direction is
    along the incident light, to the global frame. This transformation is provided
    by the  direction cosine rotation matrix:}
\begin{equation}
    R(u,v,w) = R_{\rm z'} (\varphi)\cdot R_{\rm y'} (\vartheta) =\left(\begin{array}{ccr}
    \frac{uw}{\sqrt{1-w^2}}& -\frac{v}{\sqrt{1-w^2}}&u\\
     \frac{vw}{\sqrt{1-w^2}}& \frac{u}{\sqrt{1-w^2}}&v\\
     -\sqrt{1-w^2}&0&w
  \end{array}\right),
\end{equation}
where the direction cosines are defined  as $u= x/r$, $v=y/r$, and $w=z/r$, where $r = \sqrt {x^2 + y^2 + z^2}$. 
We can show that the XMCD signal for any direction ($\vartheta$, $\varphi$)  is given by
\begin{equation}
\label{eq:XMCDdirection}
    \sigma_{XMCD}(\omega)= u \sigma^{yz}(\omega) + v \sigma^{zx}(\omega) + w \sigma^{xy}(\omega), 
\end{equation}
and where  $\sigma^{\mu\nu}$ is given by
\begin{equation}
\label{eq:XASgoldenrule3}
\sigma^{\mu\nu}(\omega) = \frac{4\pi \alpha \hbar}{m_e^2 \omega} \sum_{if}
    \Im \left( \langle f |p_{\mu}| i\rangle  \langle i|p_{\nu}|f\rangle  \right) 
    \delta(\hbar \omega-\epsilon_f + \epsilon_i ).
\end{equation}
Here $\Im$ is the imaginary part, and $\mu, \nu= x,\; y,\;$ or $z$.\\

We have demonstrated 
that  the plane wave contribution to the x-ray absorption matrix elements compensates almost perfectly
the pseudo partial wave contribution. The derivation of the matrix element is shown in appendix B. As shown in 
Fig. \ref{fig:xaspw}, the effects of the plane wave contribution to 
the L$_{2,3}$ XAS and XMCD   are extremely small. This was to be expected as we are only
interested in a limited  energy range  above the Fermi energy for
the L$_{2,3}$ edges, that are primarily associated with the $3d$ part of the  eigenfunctions 
localized  within the augmentation region. 
\begin{figure}[H]
\centering{
\begin{tabular}{c}
\includegraphics[scale=0.15]{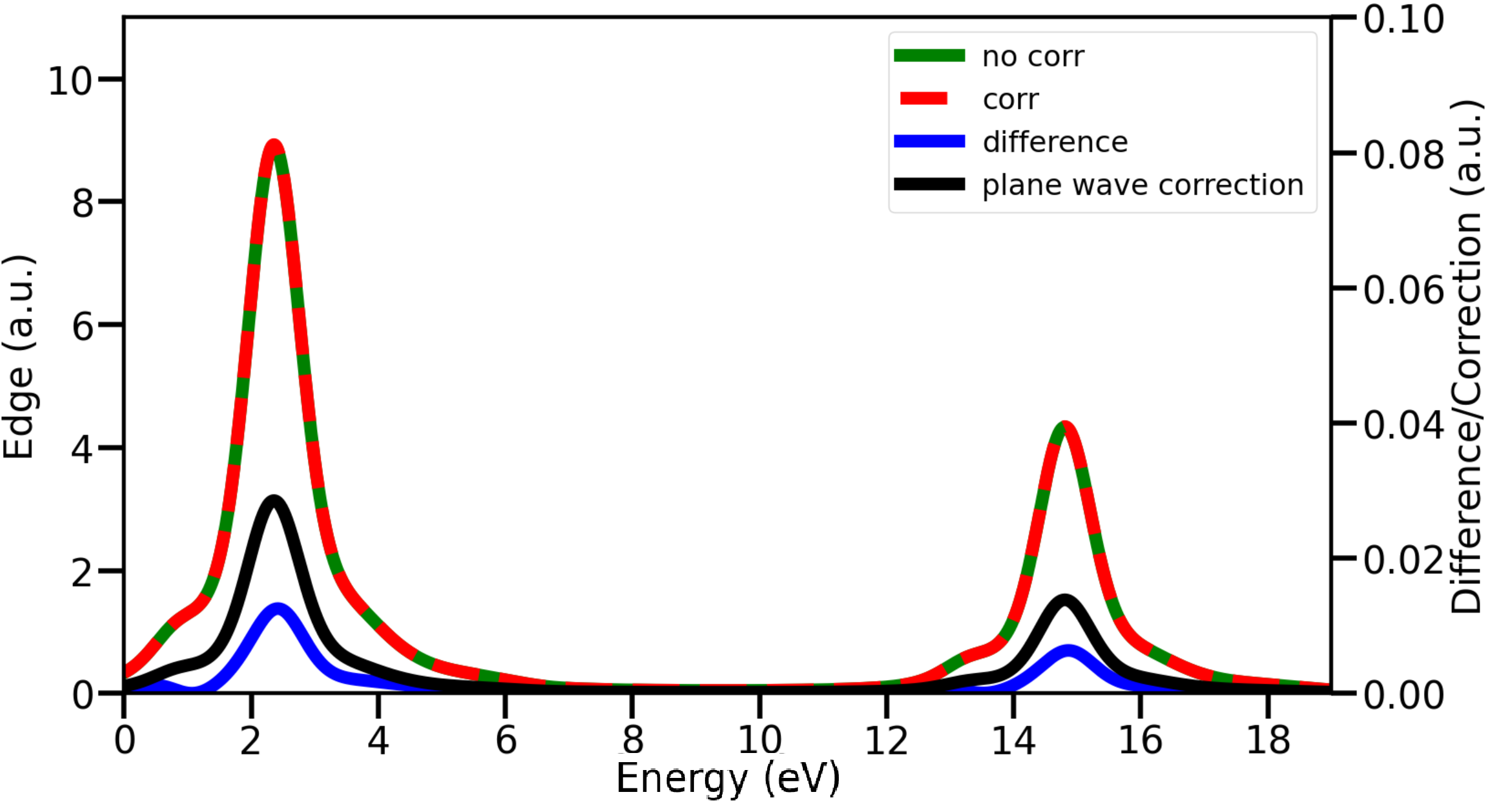}  \\
\end{tabular}
}

\caption{Comparison between the all electron partial wave (green) and the full \textsc{PAW} wave function (red) 
calculated iron L$_{2,3}$ XAS spectra for the LS state. The negligible difference (blue) of the spectra  as well
    as the plane wave contribution to the XAS (black) are shown on the right scale.}
\label{fig:xaspw}
\end{figure}
To utilize the so called XMCD sum rules\cite{Altarelli1998,PhysRevLett.68.1943,PhysRevLett.70.694} 
to compute the spin and orbital moments, we need to know to evaluate carefully the   number of electrons
in the valence states $n_e$. In principle as we are
truncating the plane wave component of the wave function, and we are therefore
restricting ourselves to the augmentation region contribution to the density of
states, which will not integrate to the theoretical values of $n_e=6$ below the
Fermi energy or $n_e=10$ over the entire energy range. In practice, doing so yields
less accurate values for the sum rules than the theoretical value of $n_h=4$,
that we will therefore use. We will also show in the results that we need to evaluate the 
magnetic dipole contribution $\langle T_z \rangle$  to obtain accurate values of the spin and orbital magnetic
moments. The $\langle T_z \rangle $  contribution will be directly  evaluated in DFT using the formula derived in the
appendix.  

\section{Results and discussion}
\begin{figure}
    \begin{subfigure}[t]{0.4\textwidth}
      \includegraphics[width=\textwidth]{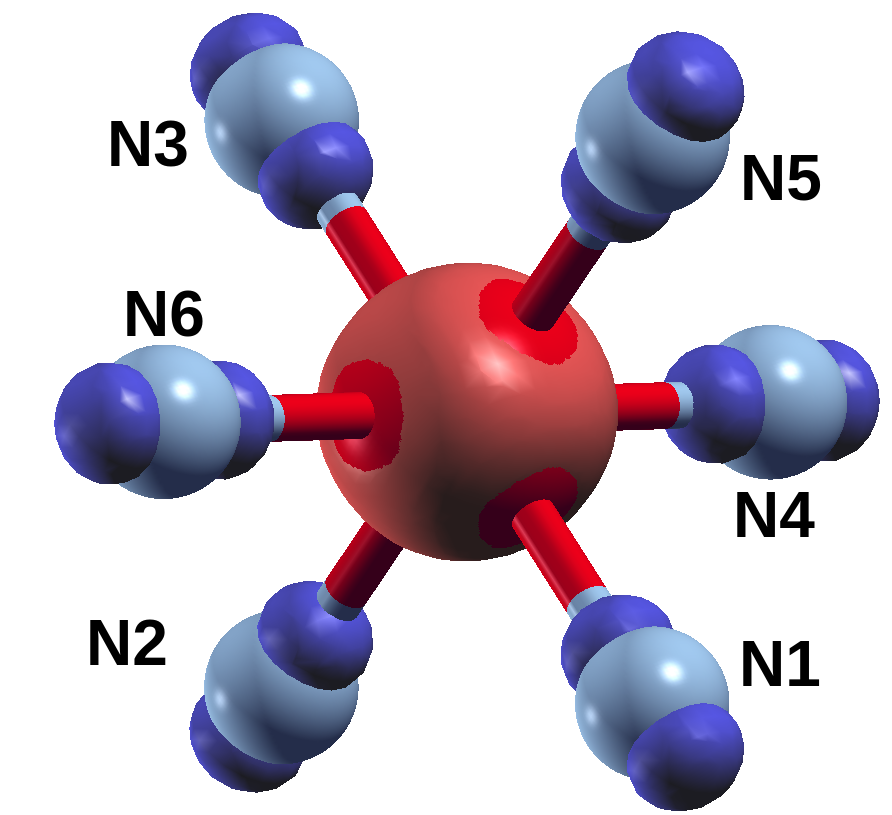}
      \caption{Undistorted FeN$_6$} 
    \end{subfigure}
    \hfill
    \begin{subfigure}[t]{0.35\textwidth}
      \includegraphics[width=\textwidth]{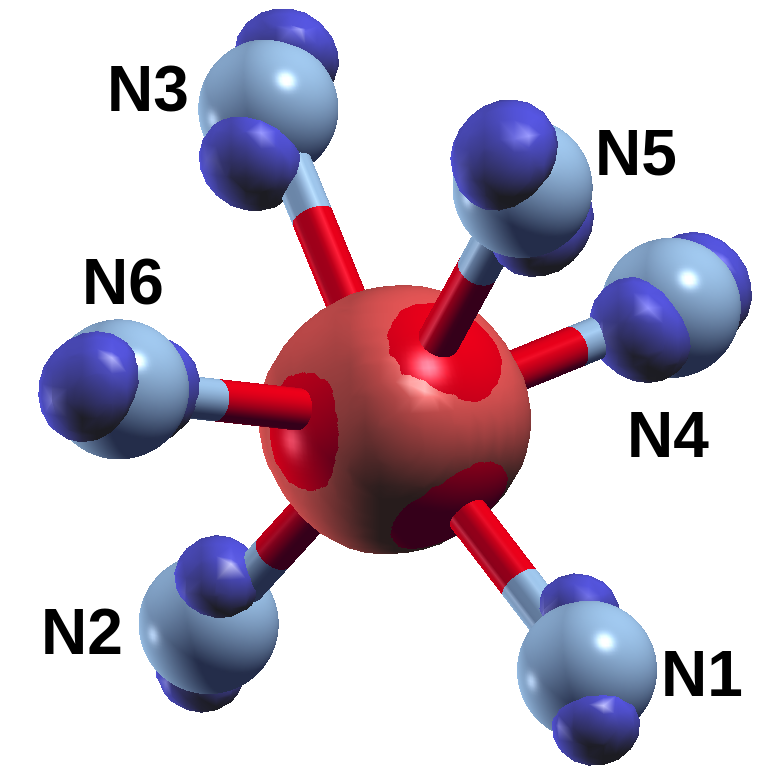}
      \caption{Distorted FeN$_6$}
    \end{subfigure}
    \begin{subfigure}[t]{0.4\textwidth}
      \includegraphics[width=\textwidth]{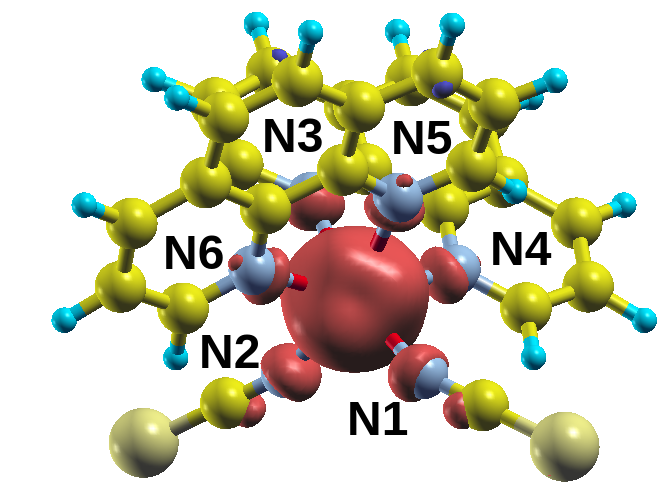}
      \caption{Fephen molecule }
    \end{subfigure}
    \caption{Comparison of the high-spin isosurface of the magnetization density of undistorted (a) 
    and distorted (b) FeN$_6$ cluster with that of the Fephen in (c) gas phase and (d) adsorbed on the Cu(001) surface.  
    The red color represents the positive magnetization (majority spin up), and blue the negative magnetization (minority  spin down). 
    For all cases, the isosurface is taken to be  $\pm 0.025 \mu_B$ per unit cell. }
\label{fig:magnetization}
  \end{figure}

To illustrate the effect of  various atoms of the molecule on the  electronic structure of the
    iron atom, and therefore on its  high spin state XAS absorption and
    XMCD, we depict in Fig. \ref{fig:magnetization} the magnetization
    isosurface at $\pm 0.025 \mu_B$ per unit cell for the undistorted (top
    left) and distorted  (top right) (as in the molecule on the copper surface)
    FeN$_6$ cluster together with that of the molecule in the gas phase (bottom
    left) and that on the substrate (bottom right). Due to the direct
    hybridization of the $p$ orbitals of nitrogen with those of the iron site, the
    magnetic moments of all the N atoms  are oriented  opposite 
    to that of the iron site. This does not apply to the case of the free
    FeN$_6$ octahedron (for more details see Supplemental Material FigS1 and FigS2
    and the effect on the iron density of states FigS3 and FigS4 \cite{SI}).  
    This is because the nitrogen atoms are chemically
    bonded to the carbon atoms of the phenanthroline.  Table
    \ref{tab:cluster}  shows the iron number of electrons and magnetic
    moment in FeN$_6$ cluster and in the molecule. It is clear
    from the table that the distortions have  only a slight impact on the electronic
    distribution of the iron atom, but as shown later, the XMCD is considerably modified.
 As a result, new $\sigma^{xz}$ and  $\sigma^{zy}$ signals
    appear. This is also true for the free molecule as compared to the molecule
    on the Cu(001) substrate, as the Fe$-$N bond lengths of the free molecule
    deviate differently from the average bond length than for the adsorbed
    molecule. {We have found that the relative root mean-square deviation (RMSD) percentage, which is
    defined as the RMSD divided by the average bond length,for the
    HS free molecule is  5.1$\%$ whereas  it is only 3.6$\%$ for the HS adsorbed molecule, and
    where the average bond length are respectively 2.15 \AA\; and 2.17 \AA. Note that for
    the low spin the respective relative RMSDs are about the same, 1.3\% and 1.6\%, with an almost equal average
    bond length of 1.96 \AA.}

\begin{table}[H]
 \centering
 \begin{tabular}{l l l l l l l  }
 \toprule
     System   & $n_s$ &$n_p$ &$n_d$ & $m_s$ ($\mu_B$)  & $m_p$ ($\mu_B$) &  $m_d$ ($\mu_B$) \\
 \midrule
     Ideal FeN$_6$ & 0.370 & 0.422& 5.838 & 0.039 & -0.029 & 3.237 \\
      Distorted FeN$_6$& 0.386 & 0.460 & 5.877 & 0.030 & -0.004 & 2.996  \\
     Fe in Fephen& 0.327 & 0.451 & 5.982 & 0.017 & 0.026 & 3.696 \\
     Fe in Fephen/Cu(001)& 0.304 & 0.407 & 6.007 & 0.017 & 0.022 & 3.635 \\
 \bottomrule
 \end{tabular}
 \caption{Iron site number of electrons and spin magnetic moments in units of  $\mu_B$ per $s$, $p$ and $d$ orbitals for ideal and distorted 
     FeN$_6$ cluster together with the Fephen molecule in the gas phase and the one  adsorbed on  Cu(001) surface.  }
    \label{tab:cluster}
 \end{table}

\begin{figure}[H]
\centering{
\begin{tabular}{c}
\includegraphics[scale=0.14]{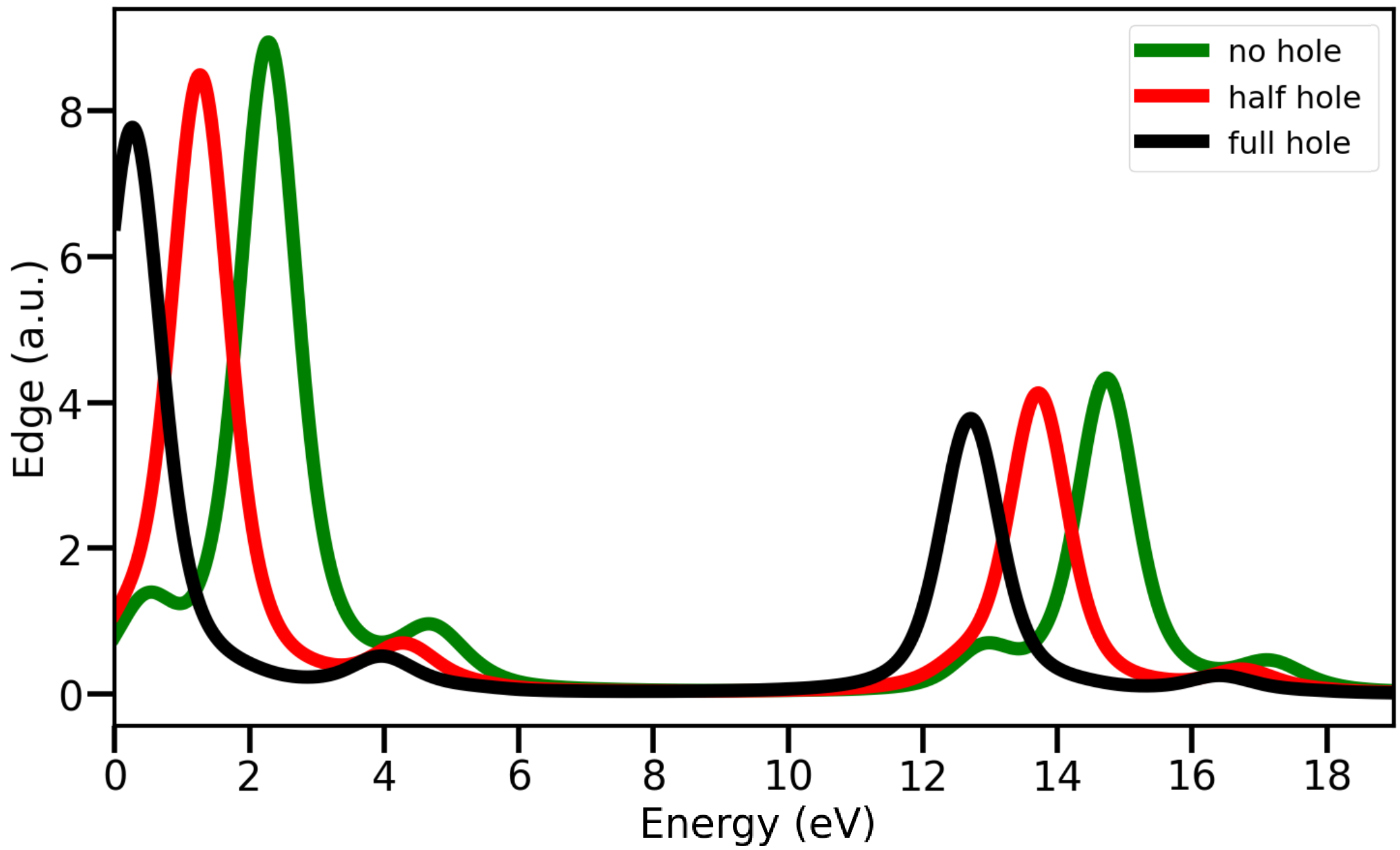} \\
\includegraphics[scale=0.14]{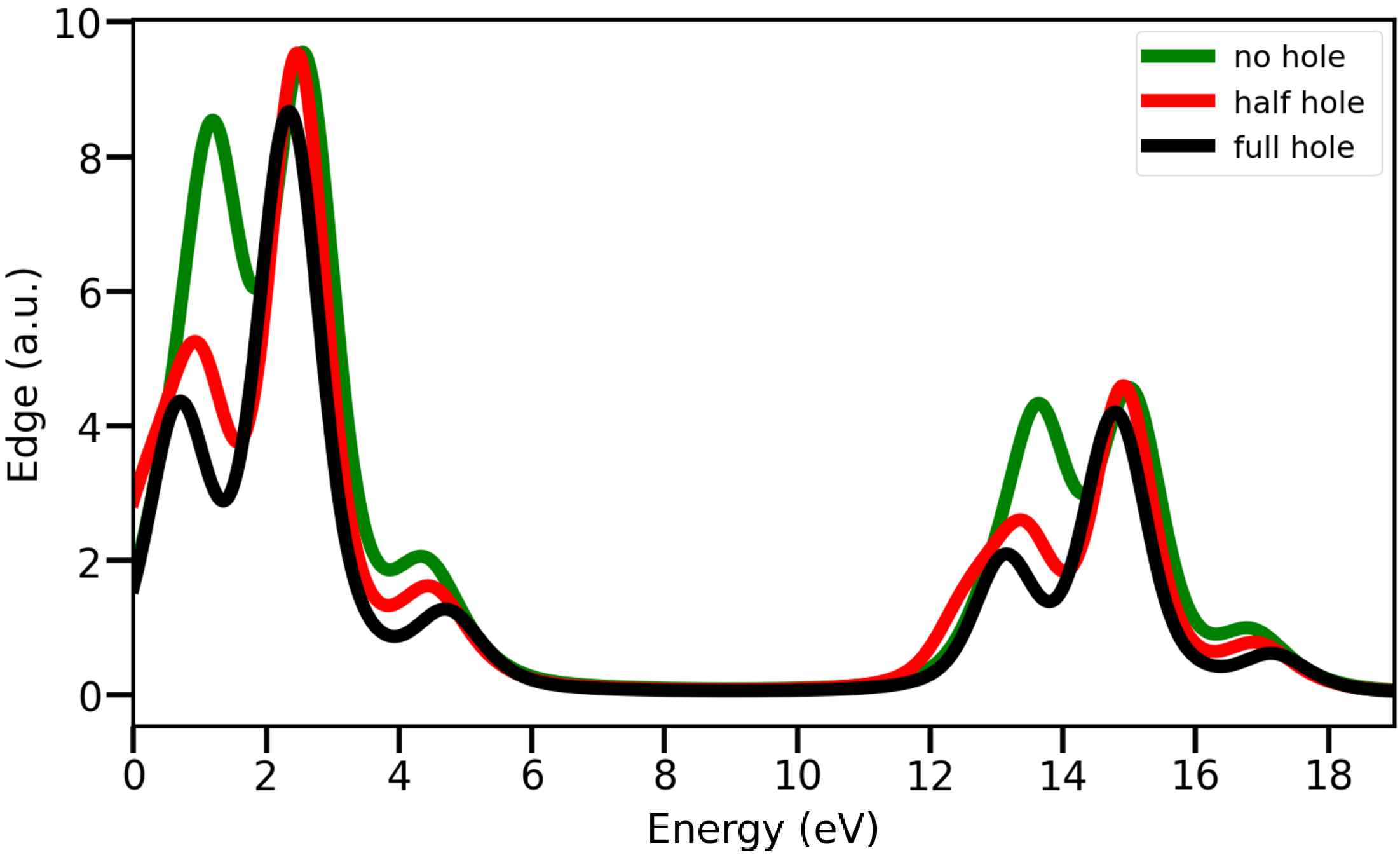} \\
\includegraphics[scale=0.39]{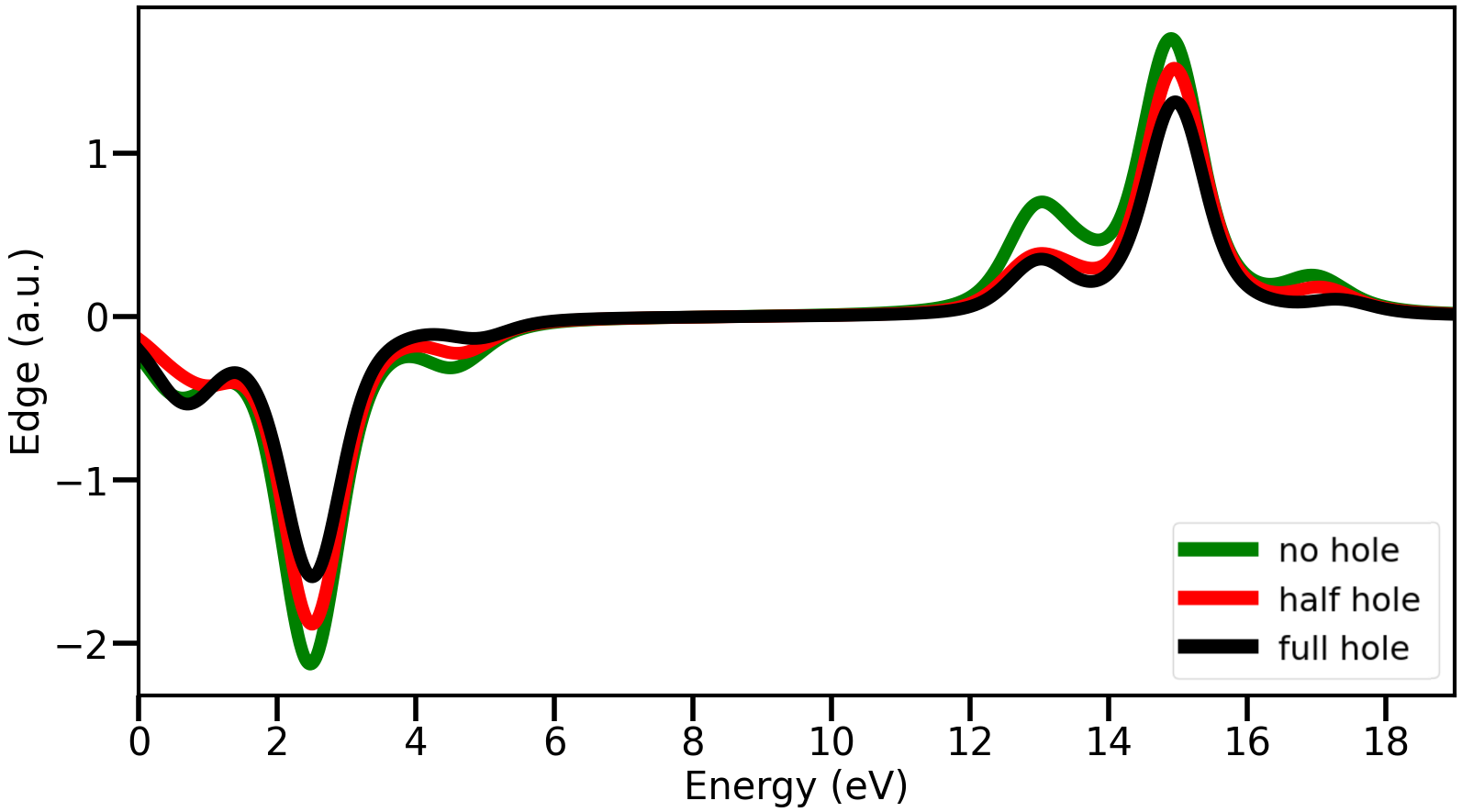} \\
\end{tabular}
}
    \caption{L$_{2,3}$ XAS for the Fephen molecule without core-hole (green), with the Slater transition rule 
    half hole (red) and the full hole (black), 
    both for the HS (top) and LS (middle) and the corresponding HS XMCD spectra (bottom).
    }
\label{fig:xas_hole}
\end{figure}

\begin{figure}[H]
\centering{
\begin{tabular}{c}
\includegraphics[scale=0.39]{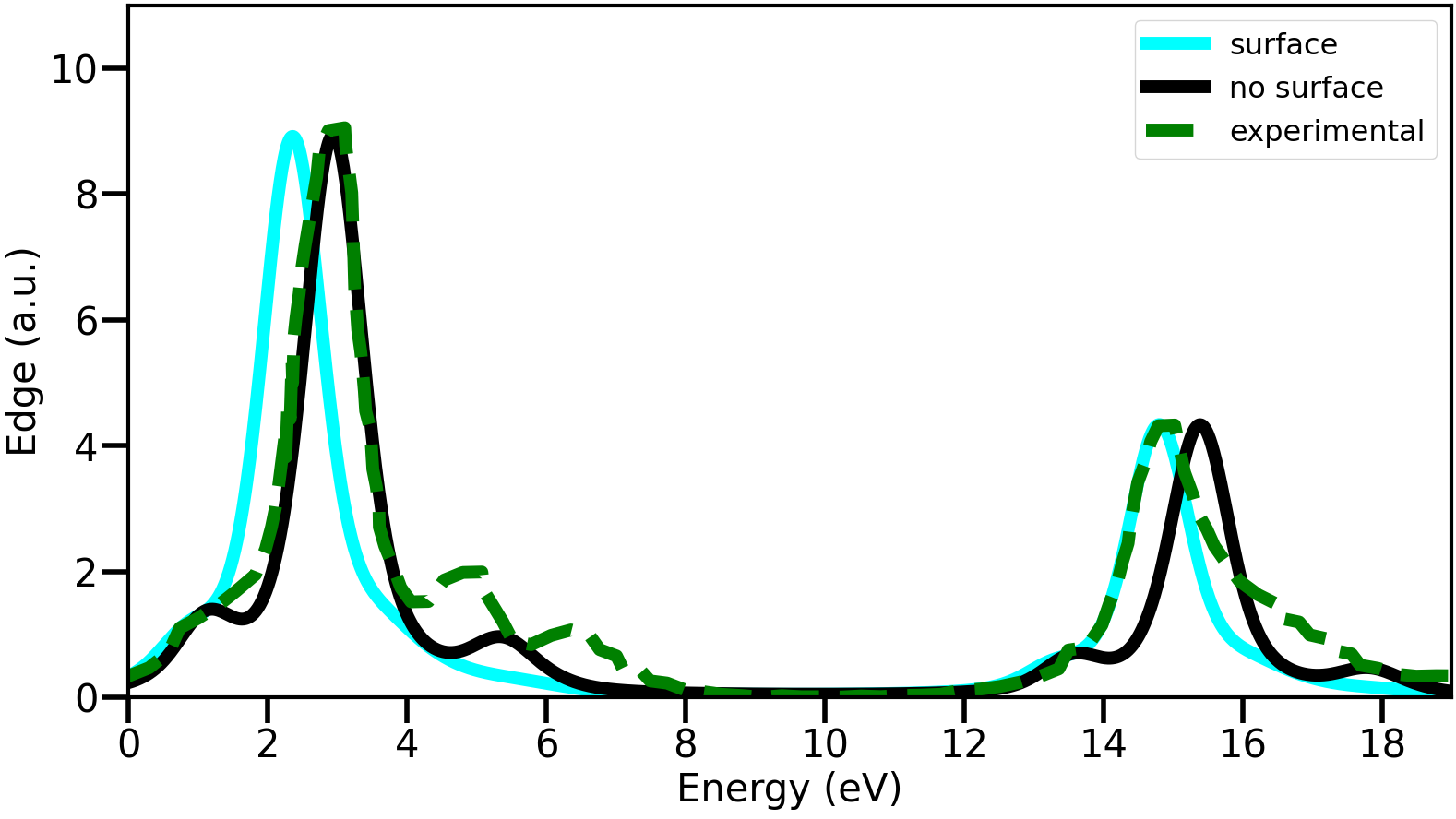} \\
\includegraphics[scale=0.39]{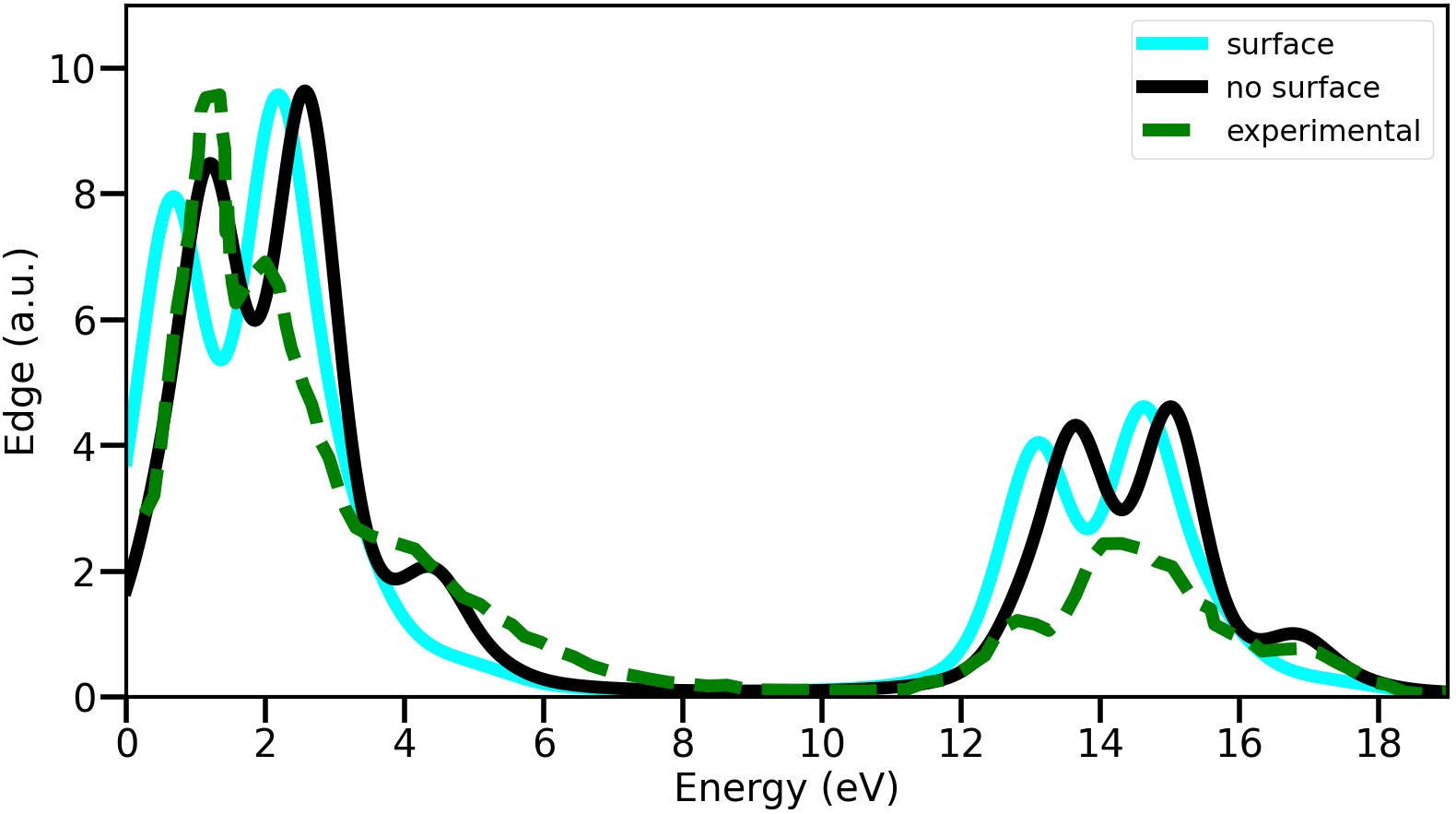} \\
\includegraphics[scale=0.39]{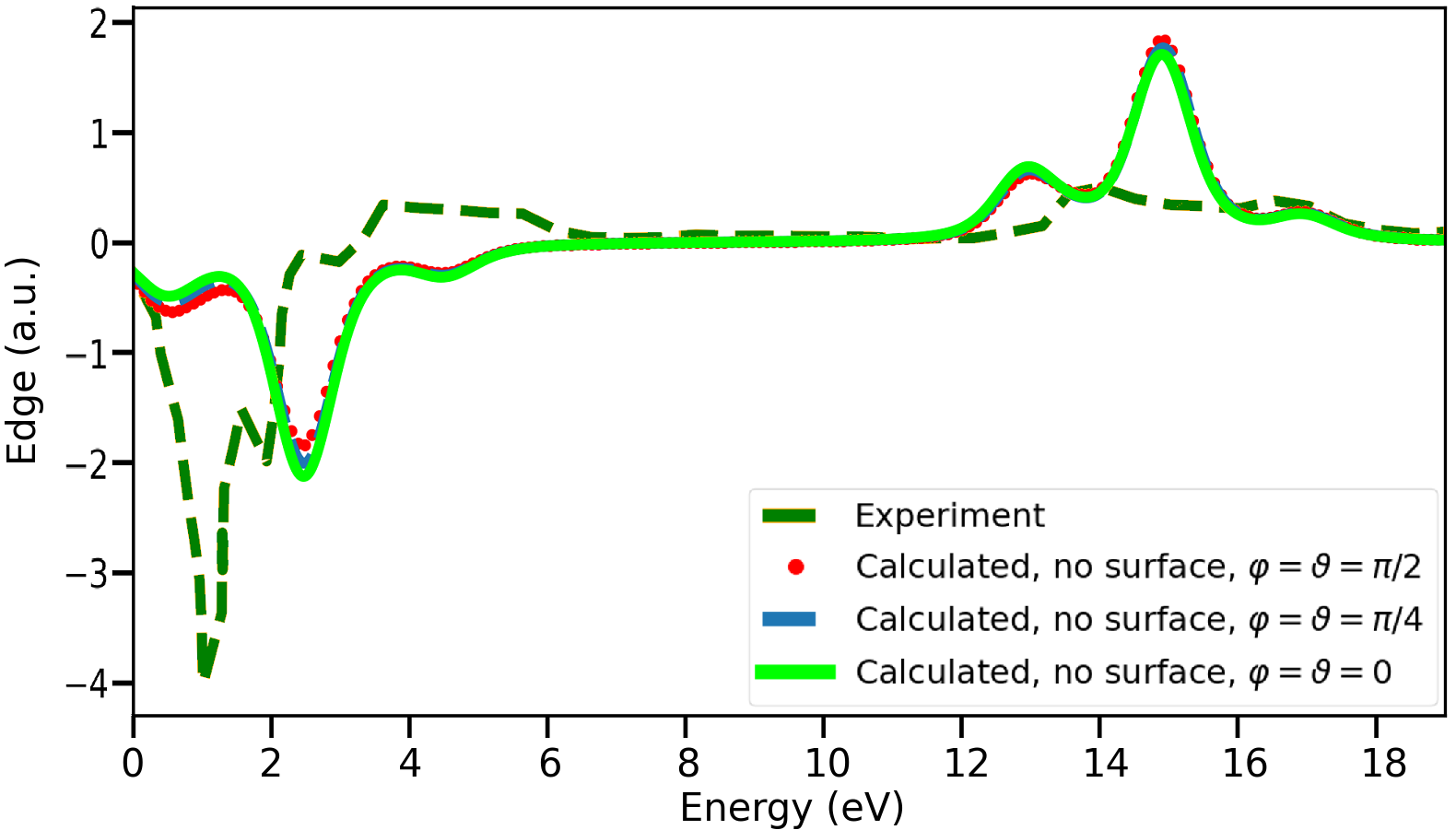} \\
\end{tabular}
}
\caption{ L$_{2,3}$ XAS for the Fephen molecule in the LS state (top), HS state 
    (second) and HS XMCD for the free molecule (third) 
    compared with the experimental results of Ref.  \cite{Miyamachi2012} for the gas phase. 
    The XMCD is calculated for the  magnetic moment aligned along the (001) direction (green), 
    the (111) (dot-dashed blue) and (010) (dashed red).
    }
\label{fig:xasexpt}
\end{figure}

{The XAS and XMCD L$_{2,3}$ spectra were computed for the Fephen molecule up to
6.5 eV above the Fermi energy. The calculations are done using the relaxed atomic positions of the molecule.
The root mean square deviation from the experimental atomic positions for the
free molecule is about 0.2 \AA\; for the HS and 0.1 \AA ~~ for the LS. The difference between the XAS
spectra calculated with the experimental positions and the calculated ones is negligible.
 The L$_2$ and L$_3$ edges are split by the
relativistic $p_{1/2}-p_{3/2}$ spin-orbit energy, which we have found to be  12.45
eV using an atomic all-electron relativistic program \cite{Koelling_1977}. The
program also produced a $p_{3/2}$ energy shift of 0.66 eV towards higher energies
for the spin-polarized state compared to the non spin-polarized one.  The
L$_{2,3}$  edges are broadened by  a Gaussian function  of full-width of   0.25
eV and a Lorentzian function of 0.5 eV, leading to a Voigt profile with a
broadening $\approx 0.6$ eV, in agreement with experimental results
\cite{CartierditMoulin1992}. To determine the effect of the static
core hole on the XAS, we have performed a calculation including a static core
hole in the core 2$p$ states, and also a half hole according to the Slater
transition rule. We have compared in Fig. \ref{fig:xas_hole} the calculation of
XAS and XMCD using the so called initial state, where no core hole is included,
with the calculation using a full core hole and that using a Slater half hole.
We observe that the structures in the  LS XAS are shifted linearly towards  lower
energies by 2 eV for the full core hole and 1 eV for the half hole. This shift
corresponds to the screening of the core hole by the additional photo-electron that
remains on the iron site.  The linear reduction of the peak intensity also
corresponds to the overall reduction in the number of unoccupied states in
the $3d$ density of states of iron. The situation is similar for the HS XAS, but
it is less pronounced than for the LS case, and we observe also a clear reduction of the peak
intensities.} 
{The reduction of the peak intensity of the first structure is
clearly in disagreement with experiment as it will be shown in Fig.
\ref{fig:xasexpt}.   The calculated spin magnetic moment is also drastically
reduced from 3.696 $\mu_B$ without core hole to 3.06 $\mu_B$ with a half core
hole and 2.452 $\mu_B$ for a full selfconsistent core hole. This shows the
limitation of XAS calculations using a static core hole. Unfortunately,
calculations beyond a static core hole, such as using the Bethe-Salpeter equation to 
compute the electron-hole
interaction, are not feasible for such a large system because of the prohibitive
computer cost\cite{PhysRevB.83.115106,PhysRevB.82.205104} and it is the reason 
that the formalism have been used only for materials with only few atoms per unit cell.}


  {Since calculations including a static core hole deviate more from the experimental XAS spectra of Miyamachi \textsl{et al.} \cite{Miyamachi2012},  we have shown in 
    Fig.  \ref{fig:xasexpt} only the calculated x-ray XAS and XMCD  using the initial state and have compared our
    results to the experimental spectra. 
    We have adjusted the energy
    reference by shifting the theoretical LS spectrum relative to  the HS
    one by our computed value of $0.66$ eV. 
    We have then plotted the experimental LS spectrum for the gas phase by adjusting the 
    well-defined L$_3$ peak as a reference point, and the HS spectrum being then automatically
    obtained and compared to the experimental one.
    We can see that the calculation  reproduces the L$_{2,3}$ edges in
    the LS state, but as expected the multiplets, which are not taken into
    account in the calculation, are not reproduced. It is interesting to note 
    that the spectrum for the molecule on the surface is similar to that of the gas phase, apart from the structure at about 5 eV which is strongly reduced. 
    The two spectra for  the gas phase  and for the adsorbed molecule are shifted by the difference of their 
    respective Fermi levels. 
    In the case of the HS
    state, we note a less accurate agreement between the theoretical and
    experimental results  as the experimental
    peaks are slightly shifted and have different intensities for the L$_2$ and L$_3$ edges. These differences
    might be dependent on the dynamics of the  core-hole\cite{PhysRevLett.80.4586}, although we still have a
    qualitative agreement. } 

    For the XMCD, we have made  calculations
    for three  alignments of the magnetic moment. The first for the moment along the (001) easy axis, 
    and the second and the 
    third for the moment along the (111) and (010) directions. 
    Note that the magnetic moment direction has a negligible effect on the total 
    XAS.  These XMCD results for various magnetization directions  will be  used later to determine 
    the orbital magnetic moment anisotropy by means of
    the XMCD sum rules.  As it can be seen from Fig. \ref{fig:xasexpt} the agreement with the experimental data
    is only qualitative. This  is  expected, as the XMCD simulation is
    notoriously complex. It relies on the difference between two relatively close
    spectra for left and right circular polarizations, and it is therefore  extremely
    sensitive to numerical errors and approximations. Indeed, one can easily
    observe that the XMCD spectrum vanishes exactly if spin-orbit coupling is not taken into
    account as both spin channels will then couple identically with the photon
    helicity, and therefore both left and right polarizations give the same
    results. As such, the value of the orbital moment is
    strongly dependent on the accuracy of the spin-orbit treatment,
    which, therefore, constitutes an important source of error as
    it is numerically very difficult to compute accurately for such a large molecule.

\begin{figure}[H]
\centering{
\begin{tabular}{c}
\includegraphics[scale=0.39]{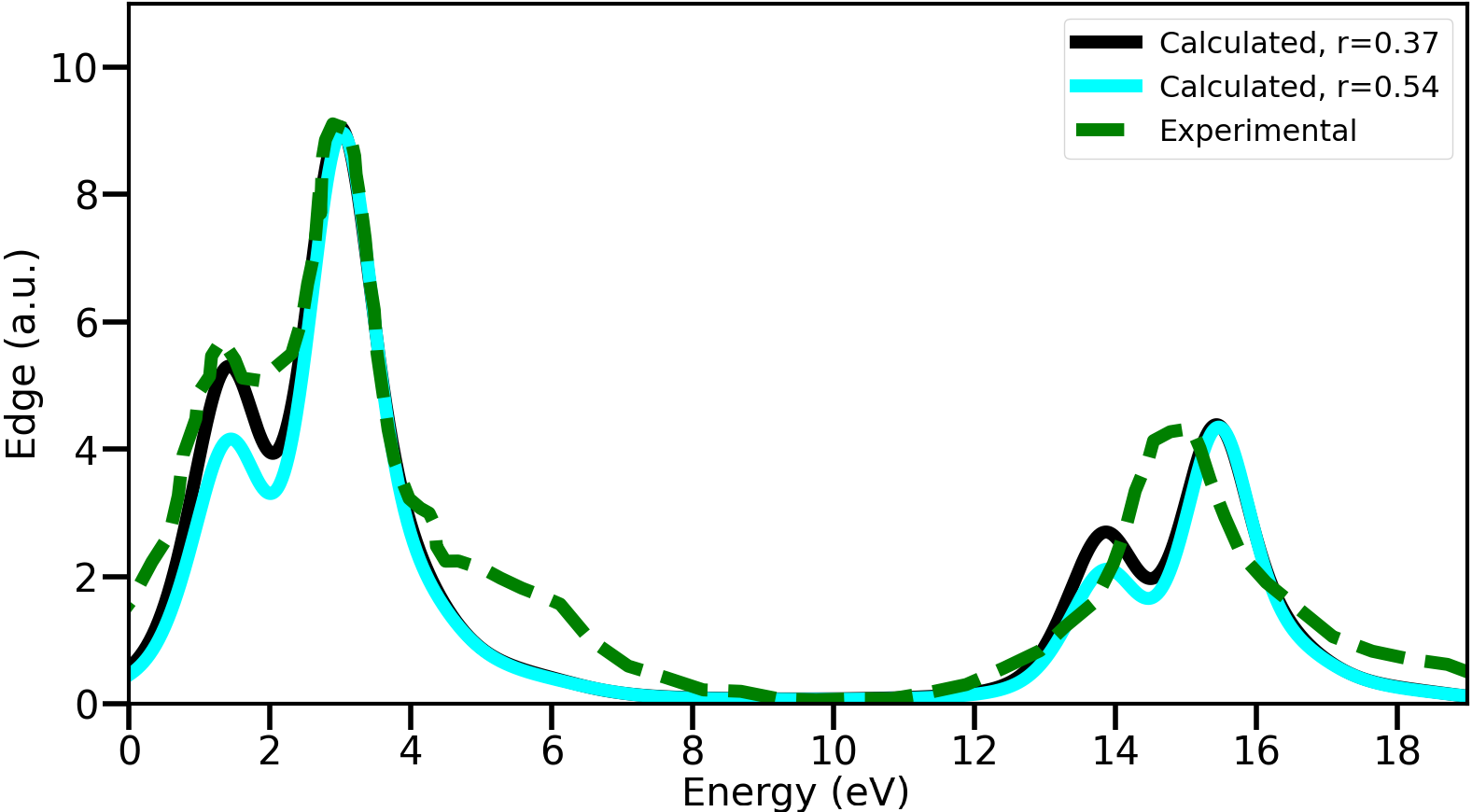} \\
\includegraphics[scale=0.39]{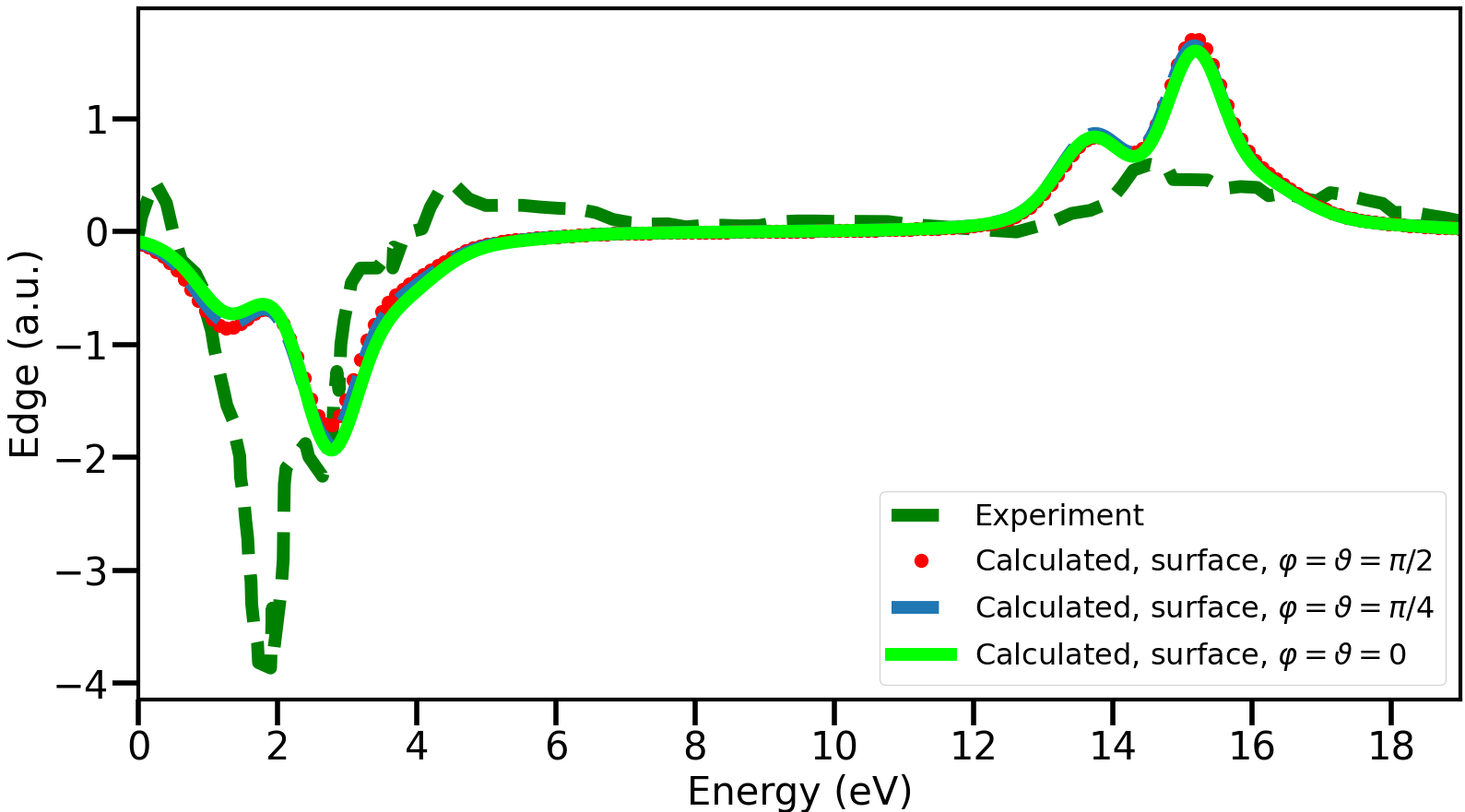} \\
\end{tabular}
}
    \caption{Linear combination of  LS (37\% black, 54\% blue)   and HS (63\% black, 46\% blue) 
    L$_{2,3}$ XAS for the Fephen molecule on Cu(001) 
compared with the experimental results of Ref.  \cite{Miyamachi2012}. The bottom figure shows the
    XMCD at the L$_{2,3}$ of iron for the molecule adsorbed on Cu(001) compared to the experimental 
    results of  \cite{Miyamachi2012}. The XMCD is calculated for the  magnetic moment aligned along the 
    (001) direction (green), the (111) (dot-dashed blue) and (010) (dashed red).
    }
\label{fig:xasexptsurf}
\end{figure}

{So far we have only compared our calculation to the experimental results of the
gas phase Fephen molecule. However, Miyamachi \textsl{et
al.}\cite{Miyamachi2012} have also measured the XAS of two layers of Fephen molecules
adsorbed on the Cu(001) surface and found that the L$_{2,3}$ XAS corresponds to
a mixture of 46\%\; of  HS and 54\%\; of  LS signals.
The LS and HS mixture is
found by fitting the XAS of the two layers of Fephen molecules adsorbed on
Cu(001) using a linear combination of the powder XAS spectra of the LS and HS. 
It is surprising to
notice that the XAS for powder  provided a good fit to the XAS of the
molecule adsorbed on the Cu(001) surface. To shed some light on the
experimental data, we have also made a fit of the experimental XAS by combining
our LS and HS XAS of the molecule adsorbed on the surface. We have proceeded as
in the  experiment by linearly combining our spectra: $r \sigma^{LS}(\omega+\Delta) + (1-r)
\sigma^{HS}(\omega+\Delta) $, where $\Delta = 0.7$ eV is the energy difference
between the corresponding peak positions of the LS  XAS of the gas phase and
that of the molecule adsorbed on the Cu(001) surface, and $r$ is the amount of
LS proportion. 

As shown in Fig. \ref{fig:xasexptsurf} (top) the amount $r$ of the LS  is found to
be 37\% for the  best agreement with experiment, whereas experimentalists have found a value of 54\%.  We have also shown the  spectrum corresponding to the
experimental  LS amount of 54\%  which was not very  different from that of the optimal
$r$ amount. As stated above the agreement of the experimental XAS
spectra of the gas phase molecule with the one where the molecule is adsorbed
on the surface is surprising. As shown by our calculation, the Fermi level of
the molecule adsorbed on the surface is shifted by 1.2 eV towards higher energy
with respect to the gas phase. It is therefore unlikely that the XAS
experimental peak positions of the free molecule and the one adsorbed on the
surface are not shifted with respect to one another, but it is also possible that a
significant amount  of the molecules is not adsorbed on the surface as the
Fephen had a 2 ML thickness. This claim could be experimentally verified
and explored, e.g. by  measuring the spectra of one Fephen monolayer or less  on the surface.  
Fig.\ref{fig:xasexptsurf} bottom shows also the calculated and measured XMCD. The
agreement with experiment is similar to that for the molecule on the gas phase.}

\begin{figure}[H]
\centering{
\begin{tabular}{c}
\includegraphics[scale=0.15]{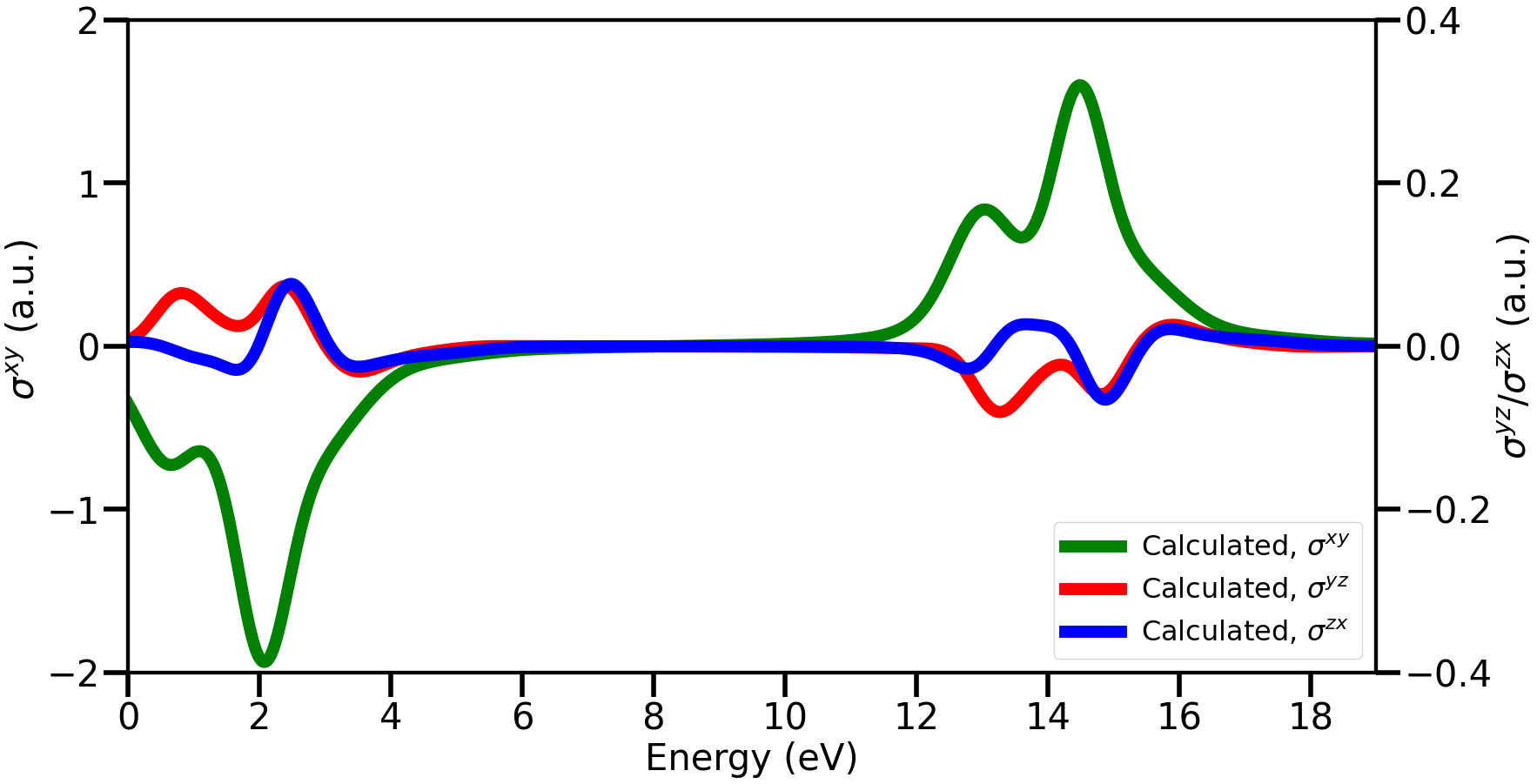} \\
\includegraphics[scale=0.15]{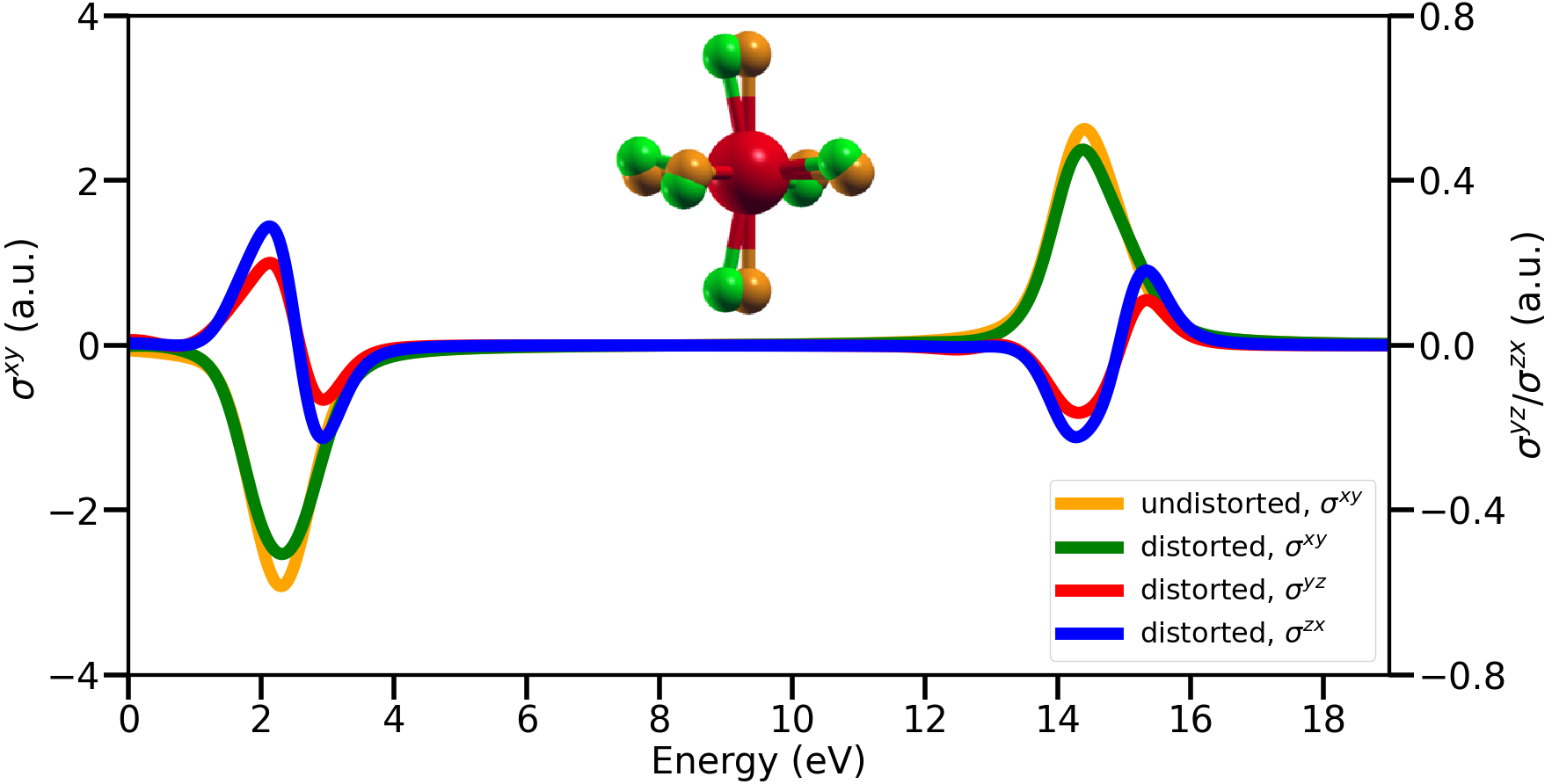} \\
\end{tabular}
}
    \caption{ Calculated $\sigma^{\mu\nu}$ (see Eq. \ref{eq:XASgoldenrule3})  for the HS Fephen molecule (top) when the
    magnetic moment is aligned along the (001) axis and for perfect and deformed FeN$_6$ octahedron (bottom). $\sigma^{yz}$ 
    and $\sigma^{zx}$  are strictly zero by symmetry for the undistorted octahedron (not shown). The scale of $\sigma^{yz}$ 
    and $\sigma^{zx}$ is on the right.   The inset shows the
    the deformation of the octahedron (Nitrogen atoms in green)  as in the molecule case compared to the non 
    deformed one (Nitrogen in orange). 
    }
\label{fig:distorsion}
\end{figure}

    As explained in the method of calculation, we have found that the
    dependance of the XMCD  on the direction of the incident light can be used to provide the deformation
    of the iron octahedron, whereas the dependance on the magnetization direction
    produces the anisotropy of the orbital magnetic moment.
    To support our idea, we have
    depicted in Fig. \ref{fig:distorsion} the $\sigma^{\mu\nu}$ components as given by Eq. \ref{eq:XASgoldenrule3} for 
    the $L_{2,3}$ iron atom in the molecule on Cu(001) and for the iron atom in the distorted and  undistorted  
    FeN$_6$ octahedra. For the undistorted tetrahedron, the $\sigma^{yz}$ and $\sigma^{zx}$ are exactly zero due to  symmetry
    and the code also produces zero, whereas these two components of the $\sigma$ tensor 
    do not vanish for the distorted octahedron, as shown in  Fig. \ref{fig:distorsion} (bottom). It is clear that if we set the direction of 
    the circularly polarized light along (010) or (100) direction, while keeping the magnetization along the (001) direction,
    we will observe only $\sigma^{yz}$ or $\sigma^{zx}$ as shown by Eq.  \ref{eq:XASgoldenrule3}. It is therefore interesting
    to emphasize that this kind of experiment will directly give the effect of the octahedron distortion on the XMCD signal. 
    One can set a database of XMCD spectra for  a direction where the 
    the XMCD should be zero for a perfect octahedron and machine learning can be used to predict the octahedron
    distortion of SCO molecules adsorbed on metallic surfaces.
    These theoretical predictions are interesting and need future
    experimental confirmation, as most available results yet
    deal with the crystalline phase which is clearly isotropic.

\begin{figure}[H]
\centering{
\begin{tabular}{c}
\includegraphics[scale=0.15]{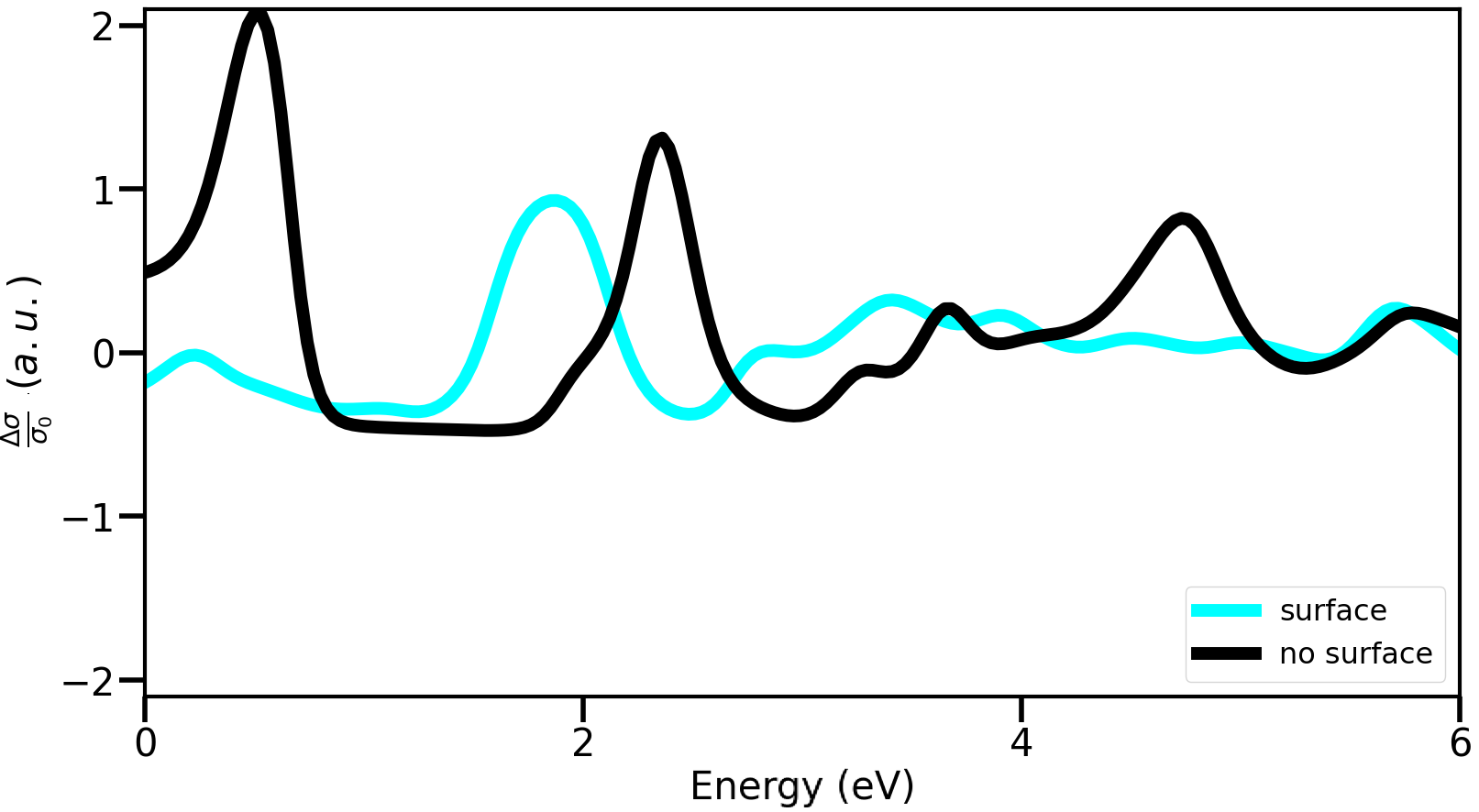} \\
\end{tabular}
}
    \caption{Calculated XAS anisotropy  $\Delta\sigma /\sigma^{0}$, where $\Delta\sigma = (\sigma^+ +\sigma^-)/2 - \sigma^0$  
    for the molecule in the gas phase (black) and adsorbed on Cu(001) surface (blue). 
    }
\label{fig:anisotropy}
\end{figure}

{To describe further the distortion from the octahedral symmetry on the x-ray
absorption we have first analyzed the deformation of the octahedron for both
the gas phase molecule and the one on the surface. The method used to compute
the deformation is well described in appendix A.  We have calculated the
distortion of the octahedron from an ideal one and found that in the HS the
octahedron of the free molecule is more deformed than that of the adsorbed one
(see FigS5 for further details on the effect of the  octaheron distortion on the iron L$_{2,3}$  \cite{SI}).
Indeed  the cost function defined by Eq. \ref{eq:cost} has a value of   0.016
in the gas phase but only 0.011 in for adsorbed  molecule as shown in  table IV
of appendix A. This is also compatible with the relative RMSD of the bond length discussed previously. 
We have therefore analyzed the anisotropy of the x-ray
absorption by computing $\Delta\sigma / \sigma^{0}$, where $\Delta\sigma =
(\sigma^+ +\sigma^-)/2 - \sigma^0$, as shown in Fig. \ref{fig:anisotropy}.
Note that this anisotropy should be  zero for a perfect octahedron and   a non
zero value  gives us the degree of deformation of the octahedron.  As is
expected, the figure shows that this anisotropy is strong, and it is much larger
for the  free molecule compared to the adsorbed molecule.  This is unexpected
as we might assume that the surface will deform further the octahedron. In fact
the opposite happens when the molecule is adsorbed on the surface because the
octahedral angle between the NCS groups is reduced from 102.4  to 96.3 degrees.
This reduction is certainly due to the lattice spacing between the surface
copper atoms which constrains the sulfur-sulfur distance (see Fig.
\ref{fig:polarization} as the  sulfur  of the NCS group is known to establish
strong bonds with transition metals.   }

%
%

\begin{figure}[H]
\centering{
\begin{tabular}{c}
\includegraphics[scale=0.12]{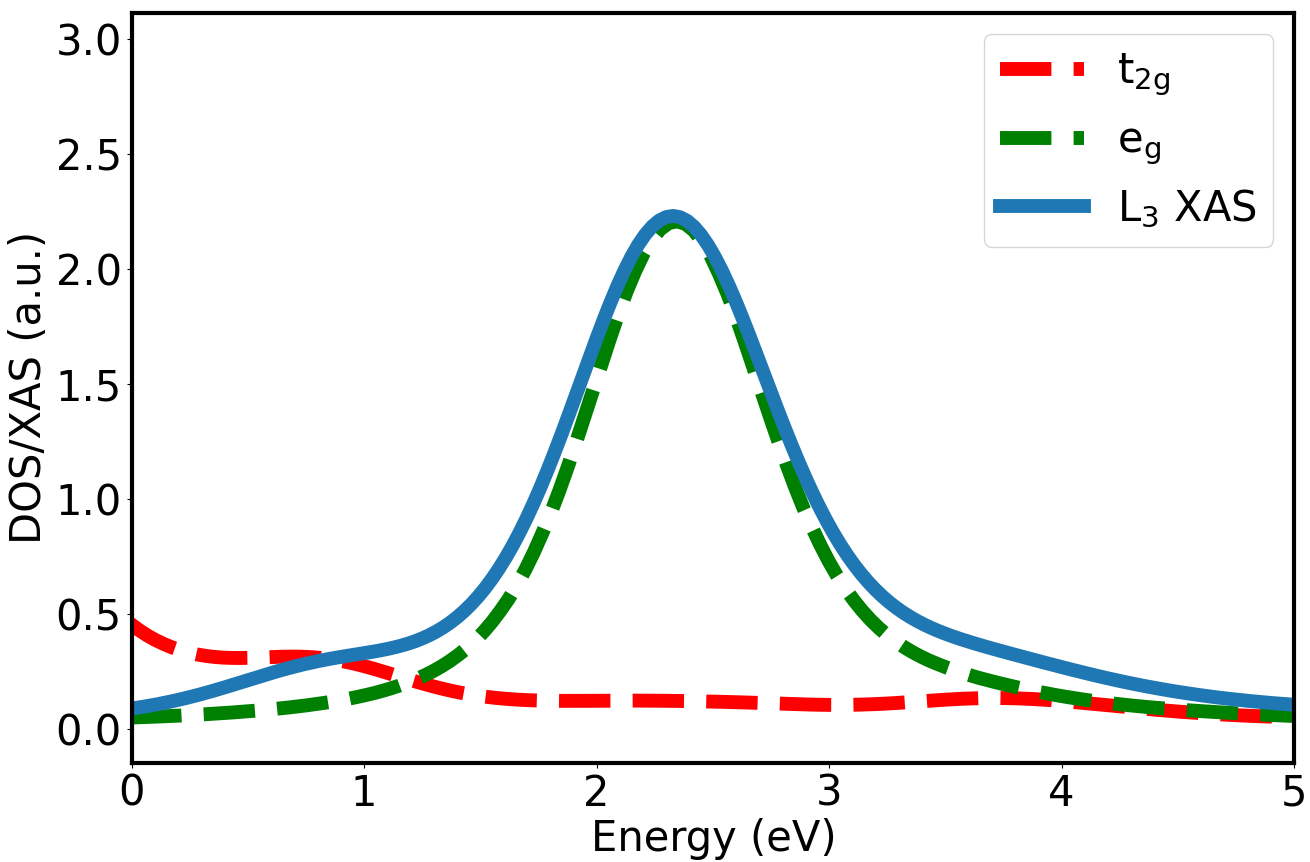}  \\
\includegraphics[scale=0.1]{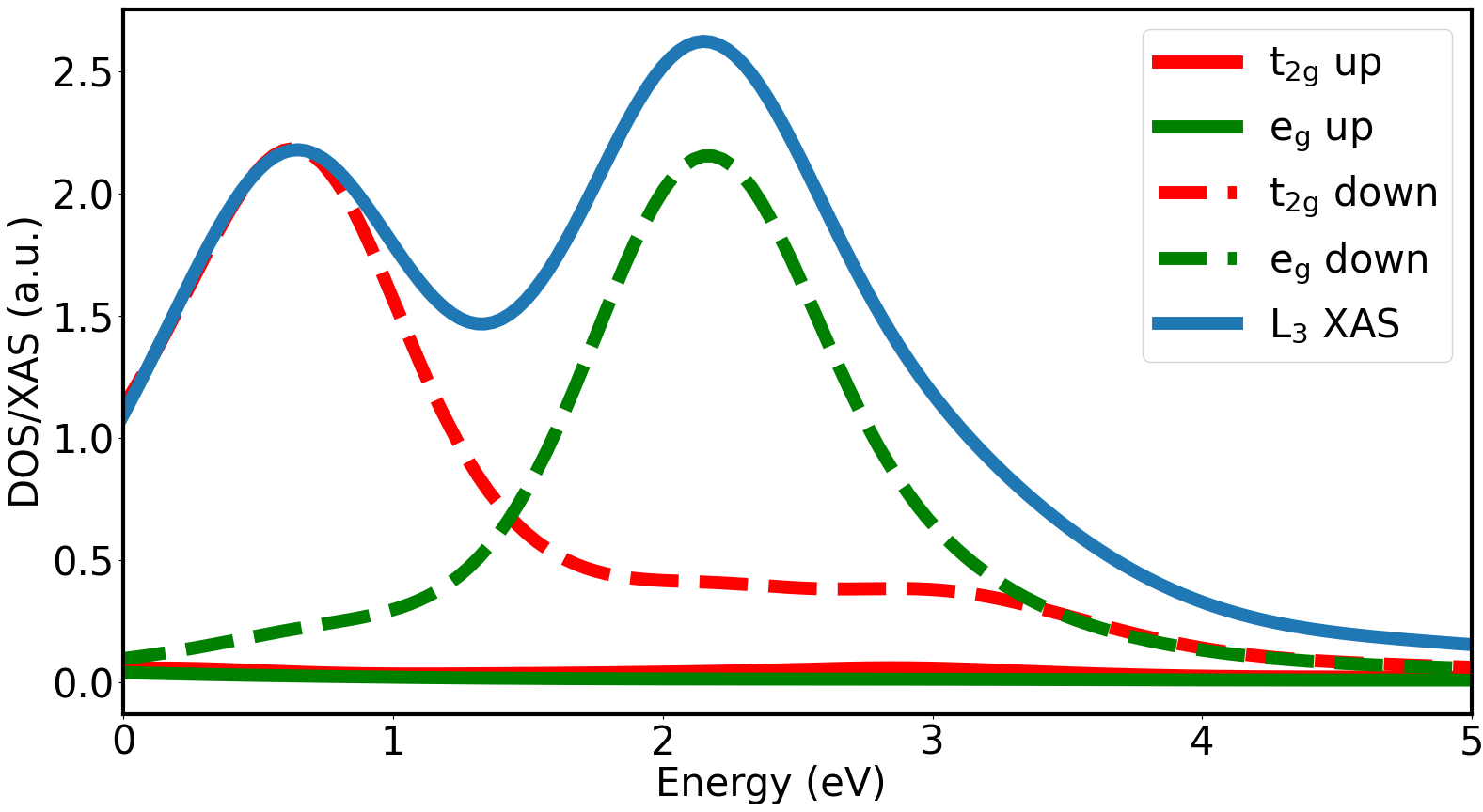}  \\
\includegraphics[scale=0.1]{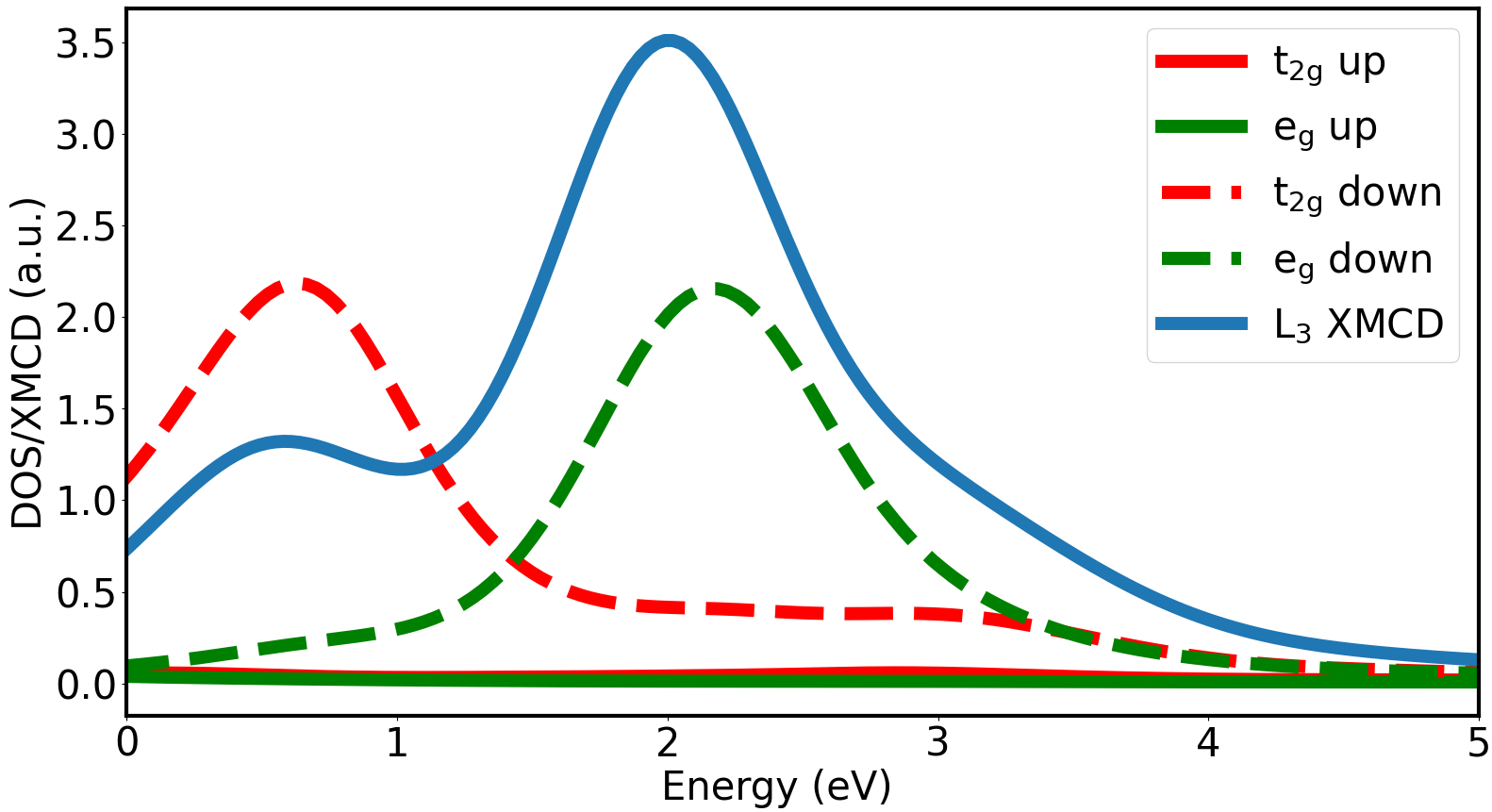}  \\
\end{tabular}
}

\caption{ The iron site spin polarized
    symmetry broken $e_g$  and $t_{2g}$ decomposed unoccupied density of states compared to the calculated XAS L$_3$ spectrum 
    for both LS (top) and HS (middle) and to HS XMCD (bottom). Note the XMCD sign is flipped for an easy comparison.}
    \label{fig:xasdos}
\end{figure}

To understand the structures in the XAS, we have compared them in Fig.
\ref{fig:xasdos} with the spin-polarized symmetry broken $e_g$ and $t_{2g}$
representations of the iron site density of states (for more details see Supplemental Material FigS3 and FigS4 \cite{SI}).  
As expected from
the $d^6$ electronic configuration  of the ground state, the density of states
shows that the primary states contributing to the LS XAS  signal are from the
unoccupied parent $e_g$ states.  As for the HS XAS,  the main
contribution are from the minority spin parent splitted $t_{2g}$ and $e_g$ states. 
These splittings of the $e_g$ and $t_{2g}$ states  are due to  
both the strong crystal-field effect and the distorted iron octahedron.\cite{Pasquier2022} 
This interpretation is compatible with the  structural and electronic structure transition from the HS to LS
which involves the spin transition 
$ (t_{2g})^{3\uparrow} (e_g)^{2\uparrow}  
(t_{2g})^{1\downarrow}  \rightarrow (t_{2g})^{3\uparrow} (t_{2g})^{3\downarrow}  (e_g)^{0} $ 
as shown in Ref. \cite{Pasquier2022}.
The figure also shows that the states contributing to the HS XMCD
spectrum are naturally the same as those for the HS XAS.

It is evident that the sum rules should be vanishing in the LS state and the numerical
calculation concurs with this analytical result.
However, this is not the case in the HS state, for which we get the
results shown in Table \ref{tab:sumrule}. These results are obtained from the XMCD spectra presented in Fig. \ref{fig:xasexpt} 
when the polarization is along the (001), (111) and (010) directions. 
We did not evaluate the sum rules for the experimental spectrum because we encountered normalization issues, which lead to 
    nonphysical values (e.g. $m_\ell$ has a computed value of several $\mu_B$). The table also shows  that the  magneto-crystalline
    energy $\Delta E$ is  lowest  when the magnetic moment is oriented along the (001) direction and the hard axis is  aligned with the (010) direction. 
    This energy $\Delta E$ is  equivalent  to  a blocking temperature of  22 K, which is very low considering the super-paramagnetic aspect of the molecular crystal. 
\setlength{\tabcolsep}{1em}
\begin{table}[H]
\centering
\begin{tabular}{c c @{\hskip 0.3in}c@{\hskip 0.3in}cc@{\hskip 0.3in}cc}
\toprule
    & $\Delta E$ (meV) & $T_z$ ($\mu_B$) & \multicolumn{2}{c@{\hskip 0.6in}}{$m_\ell$ ($\mu_B$)} & \multicolumn{2}{@{\hskip 0.0in}c}{$m_s$ ($\mu_B$)} \\
\cmidrule(lr{0.55in}){4-5}
\cmidrule(lr{0.20in}){6-7}
    Magnetization& & &Direct  & Sum rule& Direct  & Sum rule \\
direction &&&&&& \\
\midrule
    (001) & -1.9 &-0.101 & 0.180 & 0.154 & 3.635 & 3.731 \\
  (111) & -0.98 &-0.092 & 0.106 & 0.092 & 3.637 & 3.735\\
    (010) & 0 &  -0.092 & 0.006 & 0.006  & 3.635 & 3.734 \\
\bottomrule
\end{tabular}
    \caption{Magneto-crystalline anisotropy energy $\Delta E$ in meV, direct calculation of magnetic dipole ($T_z$), 
    spin moment ($m_s$), and orbital moment ($m_\ell$) compared to those
    obtained using the XMCD sum rules (in units of Bohr magneton $\mu_B$) for the iron site of the SCO Fephen in the HS state for 
    various directions of the iron magnetic moment.  The hard axis total energy is -1045.48985 eV.  }
    \label{tab:sumrule}
\end{table}
We can make several interesting observations: 
\begin{itemize} \item The magnetic dipole tensor is
            non-vanishing, and its value is non-negligible as it   makes up for
            roughly $10\%$ of the magnetic moment contribution. One could have
            expected this, as it was shown that this operator takes a finite
            value for ideal Fe$^{+2}$ octahedral complexes in the HS state when
            spin-orbit coupling is taken into account \cite{articleCroTz}. However, as we show in the
            appendices, relativistic corrections are not the
            dominant contribution to the value of this tensor in our case as they
            are almost negligible. Instead, it appears that the likely 
            origin of this behavior is imperfect
            octahedral geometry of the high spin complexes which significantly lifts
            the degeneracy of the parent $t_{2g}$ and $e_g$ states, and as
            such removes the symmetries that nullify the value of $T_z$.
    
    \item The orbital momentum sum rule yields a slightly underestimated  value of 
        $0.154 \mu_B$ compared to the directly calculated  value of $0.180 \mu_B$ when the 
        magnetization is oriented along the easy axis (001). As the accuracy of this value
        depends directly on the integral of the XMCD signal, such an agreement
        is quite surprising as one would expect a worse accordance especially
        given the relatively poor agreement between the theoretical and
        experimental spectra that demonstrated the limitation  of the model. We also note that 
        the sum rules are an approximate theoretical results, and as such the range of
        their validity has been debated \cite{PhysRevLett.80.4586,PhysRevLett.73.1994}. Nevertheless,
        according to Schwitalla and Ebert \cite{PhysRevLett.80.4586},  they should be decently 
        accurate for Fe$^{2+}$ compounds. It should also be emphasized  that 
        the integrated spectra over the relevant energy range, used for the sum rules,  are 
        in general less sensitive to the details of their structures and  shapes. \cite{PhysRevLett.73.1994}

    \item The spin-moment sum rule  appears to yield a quite accurate value of 
        $3.731 \mu_B$ compared to direct calculated value of $3.635 \mu_B$. 
        We can appreciate here
        the importance of the magnetic dipole correction, as without it the sum rule  
         would only yield a moment of $3.377 \mu_B$, which is still
        within an acceptable range from the expected value. It
        appears, therefore, that this spin sum rule is less sensitive to inaccuracies in the
        XMCD spectrum than the orbital moment sum rule. This is not surprising, as the
        orbital moment is much  smaller compared to the spin magnetic moment 
         and is consequently much more prone to errors.  

\end{itemize}

\section{Conclusion}

Our implementation of the XAS and XMCD spectra within 
\textsc{VASP} has been  used to compute the L$_{2,3}$ edges for both the low-spin and high-spin  iron site within a SCO
Fephen molecule in the gas phase and adsorbed on a copper surface. We have found that the plane wave contribution to the x-ray matrix
elements within the electric dipole approximation are small and fully compensated by the pseudo-partial contribution
to the \textsc{PAW} wave function within the XAS energy range. 

{The calculated XAS and XMCD results  are in qualitative
agreement with the available experimental results, although with relative
intensity issues in the HS state that underscore the importance of 
 multiplet and dynamics of the  core-hole effects for a comprehensive understanding of the spectrum. 
 We have found  that the simple static core hole
 or the Slater transition rule half hole did not improve the agreement with experiment due to reduction of the 
 iron magnetic moment caused by the additional screening of the core hole by the photo-electron.  The calculation using
 the initial state produced therefore the best agreement with experiment and  the overall features in both
 the HS and LS spectra 
are understood in terms of the parent symmetry broken $e_g$ and $t_{2g}$ iron $3d$ DOS. The measured XAS of Fephen
on the surface is found indeed to be a mixture of HS and LS in agreement with experiment.
We have also found that the dependence of the XMCD signal  on the direction of incident x-ray circularly polarized
light can be used to directly measure  the deformation in  the iron octahedron.
As discussed above, 
one could attempt to establish a database for the $\sigma^{yz}$ and $\sigma^{zx}$  XMCD directions, which are zeros for a 
perfect octahedron, and use machine learning to determine directly the  octahedron distortion of the SCO molecules. }

We have also shown that the XMCD for different magnetization directions
is related to the anisotropy of the iron orbital magnetic moment.
As a result,  the sum rules yield the correct orbital and spin magnetic moments as compared to direct calculations
for different orientations of the spin magnetic moment, as long as one takes into
account the contribution of the magnetic dipole moment originating from the
geometrical deformation of the iron site octahedron in the HS state.
These theoretical predictions await future experimental confirmation.

\begin{acknowledgements}
This work was performed using HPC resources from the Strasbourg Mesocenter and from the GENCI-CINES Grant gem1100.
This work of the Interdisciplinary Thematic Institute QMat, as part of
the ITI 2021 2028 program of the University of Strasbourg, CNRS and
Inserm, was supported by IdEx Unistra (ANR 10 IDEX 0002), and by
SFRI STRATUS project (ANR 20 SFRI 0012) and EUR QMAT ANR-17- EURE-0024 under the framework of the French Investments for the
Future Program. 
\end{acknowledgements}


\begin{thebibliography}{52}
\bibitem[{egt()}]{egt2g}
\bibinfo{note}{Note that strictly speaking the $e_g$ and $t_{2g}$ states form a
  symmetry adapted linear combination for the perfect octahedron as they form
  the basis set for the irreducible representations of the octahedron point
  group for the $d$ states. As the symmetry is lowered, for example by a
  Janh-Teller distortions, a new symmetry adapted combination will be in
  principle more appropriate. Here the bond lengths and bond angles are
  slightly changed from those of the pristine octahedron and all five $d$
  states levels are split. We have found it practical to keep using the parent $e_g$
  and $t_2g$ basis in order to see how the perturbative distortion affects the
  splittings of the now symmetry broken $e_g$ and $t_{2g}$ energy levels.}
\end{thebibliography}

\appendix

\section{The Magnetic dipole operator}

The magnetic dipole tensor can be defined as:
\begin{equation}
    \mathbf{\hat{T}}=\mathbf{\hat{S}}-3\mathbf{\hat{r}}(\mathbf{\hat{r}}\cdot\mathbf{\hat{S}})/|r|^2,
\end{equation}
where $\mathbf{\hat{S}}$ is the vector spin operator, and $\mathbf{\hat{r}}$ is the position operator. 
With a spin quantization axis along $z$, the magnetic dipole operator can then be written:
\begin{equation}
    \mathbf{\hat{T}}=\hat{S}_{z} -3\mathbf{\hat{r}}(\hat{r}_{z}\hat{S}_{z})/|r|^2.
\end{equation}
As stated earlier, one can show that the sum rules normally include a $\langle \hat{T}_z \rangle$ contribution in the 
valence shell of angular momentum $\ell$. We will now explicitly compute the value of this tensor using 
DFT in order to justify our approximation. We start by writing the $\hat{T}_z$ tensor as:
\begin{equation}
  \hat{T}_z= \hat{S}_{z}(1 -3\hat{r}^2_{z}/|r|^2)=\hat{S}_{z}(1-4\pi(\hat{Y}^0_{1})^2),
\end{equation}
where $Y^0_1$ is the spherical tensor operator associated to the spherical harmonic $Y^0_1$. 
Then, its mean value in some shell of electrons with angular momentum $\ell$ can be evaluated in the \textsc{PAW} method as:
\begin{align}
\label{TzPAW}
    \langle \hat{T}_z \rangle &=\sum_{n,\mathbf{k},s} f_{n,\mathbf{k},s}\langle n,\mathbf{k},s|\hat{T}_z|n,\mathbf{k},s\rangle  \\ \nonumber
    &=\sum_{n,\mathbf{k},s} f_{n,\mathbf{k},s} \sum_{\substack{ p,m \\ p',m'}}P^{* n,\mathbf{k},s}_{p',\ell,m'}P^{n,\mathbf{k},s}_{p,\ell,m}\langle p',\ell,m',s|\hat{T}_z|p,\ell,m,s\rangle ,
\end{align}
where we once again disregard the plane wave contribution as we are only interested in the augmentation 
region in this calculation, and we  have introduced the Fermi occupations $f_{n,\mathbf{k},s}$ so that the sum naturally only runs over the occupied states. 
Using its definition, the matrix elements of $\hat{T}_z$ in the partial wave basis can be written as:
\begin{align}
    \langle p',\ell,m',s|\hat{T}_z|p,\ell,m,s\rangle &= \langle p',\ell,m',s|\hat{S}_z(1 -3\hat{r}^2_z/|r|^2)|p,\ell,m,s\rangle  \\ \nonumber
    &=\langle p',\ell,m',s|\hat{S}_z(1-4\pi(\hat{Y}^0_1)^2)|p,\ell,m,s\rangle  \\ \nonumber
    &= m_{s} (\langle p',\ell,m,s|p,\ell,m,s\rangle \delta_{m,m'} - 4 \pi \langle p',\ell,m',s|(\hat{Y}^0_1)^2|p,\ell,m,s\rangle  ),
\end{align}
where $m_s$ is the magnetic moment. Using the definition of the spherical harmonics \cite{Edmonds+2016}, one can show that:
\begin{equation}
    (\hat{Y}^0_1)^2=\sqrt{\frac{1}{4\pi}}\hat{Y}^0_0+\sqrt{\frac{1}{5\pi}}\hat{Y}^0_2.
\end{equation}
This leads to the following:
\begin{equation}
    \langle p',\ell,m',s|\hat{T}_z|p,\ell,m,s\rangle = -4\sqrt{\frac{\pi}{5}} m_{s} \langle p',\ell,m',s|(\hat{Y}^0_2)|p,\ell,m,s\rangle .
\end{equation}
The matrix element involves an integral over three spherical harmonics $Y^0_2$, 
$Y^{m}_{l}$ and $Y^{m^\prime}_{l}$. This is known in the literature as a Gaunt coefficient \cite{Edmonds+2016}, 
and can be shown to be equal to:
\begin{equation}
\label{Tzgaunt}
    \langle p',\ell,m',s|(\hat{Y}^0_2)|p,\ell,m,s\rangle =\sqrt{\frac{5}{4\pi}}C^{\ell,m'}_{2,0,\ell,m}C^{\ell,0}_{2,0,\ell,0} (p',\ell| p,\ell)~~,
\end{equation}
where the $(p',\ell| p,\ell)$ are the radial integration as defined in Eq. \ref{integWigEck}.
By angular selection rules, we directly have that $m'=m$. Therefore, we obtain that:
\begin{equation}
\label{Tzfin}
    \langle \hat{T}_z\rangle =-2\sum_{n,\mathbf{k},s} f_{n,\mathbf{k},s} \sum_{\substack{ p,m \\ p'}}P^{* n,\mathbf{k},s}_{p',\ell,m}P^{n,\mathbf{k},s}_{p,\ell,m}m_{s}C^{\ell,m}_{2,0,\ell,m}C^{\ell,0}_{2,0,\ell,0} (p',\ell| p,\ell),
\end{equation}
which we implemented directly in \textsc{VASP}. 
\\\\
Note that this contribution always vanishes for a perfect $O_h$ symmetry complex of a $3d$ 
transition metal without spin-orbit coupling. In order to demonstrate this, 
we first note that in that case the magnetic dipole moment can be rewritten as:
\begin{align}
\label{perTz}
    \langle \hat{T}_z\rangle &=-4\sqrt{\frac{\pi}{5}}\sum_{i}  m_{s,i} \langle i|(\hat{Y}^0_2)|i\rangle  \\
    &=-2\sqrt{\frac{\pi}{5}}\sum_{u,d} (\langle u|(\hat{Y}^0_2)|u\rangle -\langle d|(\hat{Y}^0_2)|d\rangle ),
\end{align}
where the index $i$ runs over all the electrons of the ground state, and $u,d$ runs over the 
up and down populations respectively. For a $d^6$ Fe$^{2+}$ complex, we now 
need to consider the ground state configuration in both spin states:

\begin{itemize}
    \item In the LS state, the ground state corresponds to a closed $t_{2g}$ subshell. 
       In that case, the magnetic dipole operator is trivially vanishing as the two spin contributions that 
        are summed over are the same up to the spin sign.
    
    \item In the HS state, the ground state can be constructed by half-filling 
        all five $d$ orbitals with the same spin direction, then filling with an equal probability one 
        of the three $t_{2g}$ orbitals with an electron of opposite spin. Sum rules over the Clebsch-Gordan
        coefficients can be used to show that:
    \begin{equation}
    \label{sumperTz}
    \begin{cases}
    &\sum_{m}C^{2,m}_{2,0,2,m}=0\\
    & 2C^{2,-1}_{2,0,2,-1}+2C^{2,1}_{2,0,2,1}+C^{2,2}_{2,0,2,2}+C^{2,-2}_{2,0,2,-2}=0.
    \end{cases}
    \end{equation}
    Using the definition of the $d$ orbitals and equation (\ref{Tzgaunt}), it
        can easily be seen that the majority spin contribution to the magnetic
        dipole operator is proportional to the first line, whereas the minority
        spin is proportional to the second line. As such, the magnetic dipole
        tensor vanishes exactly in this case.
\end{itemize}

When the spin-orbit interaction is taken into account, the moment still vanishes in the LS
state as the two spinor directions are effectively degenerate in that case and
we can therefore use a very similar reasoning than in the non-relativistic
case. The case of the HS state is much more complex, and it can be shown that
the magnetic dipole tensor takes a non-vanishing value for certain ground state
geometries, including the $d^6$ geometry of Fe$^{2+}$ \cite{articleCroTz}.  With
this in mind, we computed the value of the magnetic dipole moment in our
molecules with (SOC) and without (NSOC) the spin orbit:

\setlength{\tabcolsep}{1em}
\begin{table}[H]
\centering
\begin{tabular}{c@{\hskip 0.5in}c@{\hskip 0.5in}c}
\toprule
Molecule & NSOC & SOC \\
\midrule
LS (surface/gas) & 0 & 0 \\
HS (surface) & -0.091 & -0.101 \\
HS (gas phase) & -0.150 & -0.161 \\ 
\bottomrule
\end{tabular}
\caption{Value of $T_z$ in the molecular systems in the HS and LS state (in $\mu_B$).}
\end{table}

As expected, the magnetic dipole operator vanishes in the low spin state. For
the high spin state, we immediately note that the moment is superior in the gas
phase than on the surface, but most importantly that the operator does not
vanish even without spin-orbit, and the spin orbit contribution is minimal. To
rationalize this apparent contradiction, we need to recall that our previous
reasoning was only valid for a perfect octahedral geometry, and that the
deformation of a real complex is often non-negligible especially in the HS
state. Distortion is known to play a noticeable effect on the features of x-ray
absorption spectra (see for example Ref. \cite{Val_1999}), and therefore it is
of no surprise that it should influence the value of $T_z$ (this was already
noted, but not shown explicitly, in \cite{PhysRevB.52.12766}). Informally,
the distortion breaks the ideal symmetry between the $d$-states that is observed in
Eq.  \ref{perTz}, and as consequence the sum rules of Eq. \ref{sumperTz} are no
longer applicable. Instead, each state is now a mixture weighted by the
\textsc{PAW} projections such as in equation Eq. \ref{Tzfin}, and there is no
\textit{a priori} reason for the said sum to vanish when both spin directions are
not degenerate such as in the HS state, even in the absence of SOC.  In order
to give a better illustration of this phenomenon, we will quantify the "amount"
of deformation of these systems away from their ideal geometry. We need first
to optimally rotate and rescale our system before comparing it to a reference
geometry. This is the essence of the so-called extended orthogonal Procrustes
algorithm \cite{Schonemann1970}. As a short summary, assume a set of points
$\mathbf{u}$ and a reference set of points $\mathbf{v}$ (the molecular
octahedral coordinates and an ideal octahedron coordinates respectively in our
case).  An obvious way of defining a "distance" to quantitatively compare these
structures is to carry a root median square displacement calculation (RMSD)
between these two structures, taking into account the fact that both systems
need to be properly rescaled together to have an accurate comparison. {Then, we
can recast the associated least-square deviation problem as a search for the
ideal rotation $\mathbf{R}$ and scale factor $c$ between $\mathbf{u}$ and
$\mathbf{v}$, so that we can write the cost function associated to this RMSD
calculation as:}
\begin{equation*}\label{eq:cost}
    L(\theta,\Phi)=\frac{1}{2}||\mathbf{v}-c \mathbf{R}(\theta,\Phi)\mathbf{u}||^2,
\end{equation*}
that needs to be minimized over the set of angular variables $(\theta,\Phi)$
and $c$.  For the rotation part, the solution can be
found (\cite{Schonemann1970}) by computing the singular value decomposition of the covariance matrix
$H=\mathbf{u^T}\mathbf{v}$:
\begin{equation*}
    \mathbf{H}=\mathbf{U}\mathbf{\Sigma}\mathbf{V^{T}} \rightarrow \mathbf{R}=\mathbf{V}\mathbf{\Sigma'}\mathbf{U^{T}},
\end{equation*}
where $\mathbf{\Sigma'}$ is a $3x3$ diagonal matrix with diagonal elements
$d_1=1, d_2=1$ and $d_3=\mathrm{sign}(\mathrm{det}(\mathbf{V}\mathbf{U^{T}}))$,
which is used to enforce the positive definiteness of the determinant of the rotation
matrix, so that we always have a proper transformation. For the scale factor, 
using the definition of the matrix norm $||A||=\mathrm{Tr}(A^T A)$ in the previous 
formula for the RMSD cost function, one can show (\cite{Schonemann1970}) that the minimization yields the
following result:
\begin{equation}
    c=\frac{\mathrm{Tr}(\mathbf{u^T}\mathbf{R^T}\mathbf{v})}{\mathrm{Tr}(\mathbf{u^T}\mathbf{u})}.
\end{equation}
\\
Applying these to our case, we can obtain a quantitative estimate of the
deviation of the molecular geometries from the ideal octahedral geometry. We
also add a deviation to an ideal tetrahedral geometry by comparing it to an
imperfect tetrahedron using the 4 shortest ligand bonds in our molecular
octahedron (as the average bond length in a tetrahedral complex is shorter than
for an octahedral complex):  
\setlength{\tabcolsep}{1em}
\begin{table}[H]
\label{losstable}
\centering
\begin{tabular}{c@{\hskip 0.5in}c@{\hskip 0.2in}c@{\hskip 0.2in}c@{\hskip 0.2in}c@{\hskip 0.2in}}
\toprule
& HS (gas phase) & HS (surface) & LS (gas phase) & LS(surface)\\
\midrule
Octahedral Loss & 0.016 & 0.011 & 0.002 & 0.002 \\
Tetrahedral Loss & 0.044 & 0.048 & 0.082 & 0.147\\
Ratio Oct/Tet & 0.363 & 0.229 & 0.024 & 0.014\\
\bottomrule
\end{tabular}
\caption{Value of the Loss function in the molecular systems (in \AA$^2$) with respect to an 
    ideal octahedral and tetrahedral geometry, and ratio between the two values}
\end{table}

We can see that the octahedral RMSD is an order of magnitude higher in the high
spin state than the low spin state, and the same applies to the ratio between
the octahedral and tetrahedral RMSD. As such, not only is the geometry more
strongly deformed in the high spin state than in the low spin states, the non
negligible ratio between the octahedral and tetrahedral RMSD in the HS state
shows that the absolute deformation away from the ideal case is sizeable.
Besides, the deformation in the HS state is clearly larger in the gas phase
than when the molecule is adsorbed on the surface. It is therefore of no
surprise that the magnetic dipole operator does not vanish in this case, even
without several other spin transition compounds, and the same behavior is
observed each time. Also, early results show that the moment is a very
approximately increasing function of the RMSD. However, this approach is quite
rough as it "averages" over all the angular and length distortions and
therefore it will not be able to discriminate between the finer details that
characterize distortion, and as a consequence the exact dependence of the
magnetic dipole moment with the value of the RMSD is highly non-trivial.

\section {Plane wave contribution to XAS matrix elements}

The total plane wave contribution can be split into two parts: the pseudo
partial wave contribution $|\widetilde{p,\ell,m}\rangle $ and the actual plane wave part
$|\widetilde{n,\mathbf{k},s}\rangle $, so the Golden Rule could be written as: 
\begin{equation} \label{totcrossec}
\begin{split}
    \sigma^{\mu}(\omega) &=  \frac{4\pi \alpha \hbar}{m_e^2 \omega} 
    \sum_{M,n,\mathbf{k},s} 
    \bigg| 
    \sum_{p,\ell,m,m'} 
    C^{J,M}_{\ell',m',1/2,s}
    (\langle p,\ell,m | - \langle \widetilde{p,\ell,m}|) 
    p_{\mu} | \ell',m' \rangle 
    P^*{}^{n,\mathbf{k},s}_{p,\ell,m} \\
    &+ \sum_{m'} C^{J,M}_{\ell',m',1/2,s}
    \langle \widetilde{n,\mathbf{k},s} |p_{\mu}| \ell',m' \rangle \bigg|^2 
    \delta(\hbar \omega-\epsilon_{n \mathbf{k} s} + \epsilon_{JM} ).
\end{split}
\end{equation}
Obviously, the calculation of the pseudo partial wave contribution is identical to that of the previously calculated partial wave  part.  
On the other hand, the plane wave contribution is more involved.  One starts by the plane wave expansion:
\begin{equation}
\label{PWexp}
    \langle r|\widetilde{n,\mathbf{k},s}\rangle =\frac{1}{\sqrt{\Omega}}\sum_{\mathbf{G}}c^{n,\mathbf{k},s}_{\mathbf{G}}
    \langle r|\mathbf{k}+\mathbf{G}\rangle =\frac{1}{\sqrt{\Omega}}\sum_{\mathbf{G}}c^{n,\mathbf{k},s}_{\mathbf{G}}
    e^{i(\mathbf{k}+\mathbf{G})(\mathbf{r'}+\mathbf{\tau_{\alpha}})}.
\end{equation}
where $\mathbf{r}=\mathbf{r'}+\mathbf{\tau_{\alpha}}$ is the global electron
position, split into the nucleus position $\mathbf{\tau_{\alpha}}$ and the
local position (with respect to the nucleus) $\mathbf{r'}$. Note that the
$|\mathbf{k}+\mathbf{G}\rangle $ are the eigenfunctions of the momentum operator
$p_{\mu}$:
$p_{\mu}|\mathbf{k}+\mathbf{G}\rangle =(k_{\mu}+G_{\mu})|\mathbf{k}+\mathbf{G}\rangle $. The
plane wave expansion is normalized by the system volume $\Omega$. In this
local frame, one can then carry a partial wave expansion of the plane wave:
\begin{equation}
\label{parPWexp}
    e^{i(\mathbf{k}+\mathbf{G}) \cdot \mathbf{r'}}=4\pi\sum_{\ell,m}i^{\ell}j_\ell (\left|(\mathbf{k}+\mathbf{G})\right|
    \left|\mathbf{r'}\right|)Y^{m\;*}_{\ell}(\widehat{\mathbf{k}+\mathbf{G}})Y^{m}_{\ell}(\widehat{\mathbf{r'}}),
\end{equation}
where the $j_\ell$ are the usual spherical Bessel functions. Therefore, we can write the following:
\begin{align}
    \langle \widetilde{n,\mathbf{k},s} |p_{\mu}| \ell',m' \rangle  
    &=\frac{4 \pi}{\sqrt{\Omega}}\sum_{\mathbf{G},\ell,m}i^{-\ell} 
    (c^{n,\mathbf{k},s}_{\mathbf{G}}(k_{\mu}+G_{\mu}))^\star Y^{m}_{\ell}(\widehat{\mathbf{k}+\mathbf{G}})
    e^{-i(\mathbf{k}+\mathbf{G})\cdot \mathbf{\tau_{\alpha}}} \\ \nonumber 
    & \int \mathrm{d}r r^2 j_\ell (\left|(\mathbf{k}+\mathbf{G})\right|r)
    \phi_{\ell'}(r) \int \mathrm{d}\widehat{\mathbf{r}} Y^{m'}_{\ell'}(\widehat{\mathbf{r}}) Y^{m\;*}_{\ell}(\widehat{\mathbf{r}}).
\end{align}
Then, using the orthogonality of the spherical harmonics, we get that:
\begin{equation}
\label{PWcont}
    \langle \widetilde{n,\mathbf{k},s}|p_{\mu}|\ell',m'\rangle =\frac{4 \pi}{\sqrt{\Omega}}\sum_{\mathbf{G}}i^{-\ell'} 
    (c^{n,\mathbf{k},s}_{\mathbf{G}}(k_{\mu}+G_{\mu}))^* Y^{m'}_{\ell'}
    (\widehat{\mathbf{k}+\mathbf{G}})e^{-i(\mathbf{k}+\mathbf{G})\cdot \mathbf{\tau_{\alpha}}}\int 
    \mathrm{d}r r^2 j_{\ell'} (\left|(\mathbf{k}+\mathbf{G})\right|r) \phi_{\ell'}(r).
\end{equation}
This contribution can then be added to the absorption cross section using formula (\ref{totcrossec}).

\section{Relation between XAS and local DOS}

It is interesting to note that the XAS can be shown to be directly related to the iron 3$d$ DOS, as noted for 
example in \cite{Rooke1970-ol}. In our formulation, the relation takes a very simple form. First, we will need to use
the position representation of the transition operator. Using the Schroedinger
Equation, one can easily show that the Hamiltonian $H$ and the position
operator $r_{\mu}$ follow the commutation relation
$\left[r_{\mu},H\right]=i\hbar p_{\mu}/m_e$. This allows us to rewrite the
cross section as:

\begin{equation}
     \sigma^{\mu}(\omega) = 4\pi \alpha \hbar\omega \sum_{M,n,\mathbf{k},s}\left|
     \sum_{m'}C^{J,M}_{\ell',m',1/2,s}
     \langle n,\mathbf{k},s | r_{\mu}| \ell',m'\rangle \right |^2 
     \delta(\hbar \nonumber \omega-\epsilon_{n \mathbf{k} s} + \epsilon_{J}).
\end{equation}
Note that we used here the fine-structure degeneracy of the $\epsilon_{JM}=\epsilon_J$ 
over the set of $M$ that was not relevant thus far for this study. We can then expand the squared norm as:
\begin{align}
     \sigma^{\mu}(\omega)&= 4\pi \alpha \hbar\omega \sum_{\substack{M,m',m'' \\ 
     n,\mathbf{k},s}} C^{J,M}_{\ell',m',1/2,s}C^{J,M}_{\ell',m'',1/2,s} 
     \langle n,\mathbf{k},s |r_{\mu} | \ell',m' \rangle 
     \langle \ell',m'' |r_{\mu}| n,\mathbf{k},s \rangle  \\ \nonumber 
     &\delta(\hbar \omega-\epsilon_{n \mathbf{k} s} + \epsilon_{J} ) \\
     \nonumber &= 4\pi \alpha \sum_{\substack{M,m',m'' \\ 
     n,\mathbf{k},s}}(\epsilon_{n \mathbf{k} s} - \epsilon_{J}) 
     C^{J,M}_{\ell',m',1/2,s}C^{J,M}_{\ell',m'',1/2,s} 
     \langle 
     n,\mathbf{k},s
     |r_{\mu}|
     \ell',m'
     \rangle 
     \langle 
     \ell',m''
     |r_{\mu}|
     n,\mathbf{k},s
     \rangle  \\ \nonumber 
     &\delta(\hbar \omega-\epsilon_{n \mathbf{k} s} + \epsilon_{J} ),
\end{align}
where we rewrote $\hbar \omega$ as $\epsilon_{n \mathbf{k} s} - \epsilon_{J}$ 
thanks to the delta function as we will need it for a following approximation.
We will sum over the $\ell+1/2$ edge $\sigma^{\mu}_{\ell+1/2}(\omega)$ and 
$\ell-1/2$ edge $\sigma^{\mu}_{\ell-1/2}(\omega)$. To do so, we first need to shift 
them together as both spectra have different core energy references. 
We arbitrarily take the $\ell+1/2$ edge, and shift it 
by $\Delta \omega_J=(\epsilon_{\ell-1/2}-\epsilon_{\ell+1/2})/\hbar$. Doing so yields:
\begin{align}
    \sum_J \sigma_{J}^{\mu}(\omega) \approx 4\pi \alpha  &
    \sum_{\substack{J,M,m',m'' \\ n,\mathbf{k},s}}(\epsilon_{n \mathbf{k} s} - 
    \epsilon_{\ell-1/2}) C^{J,M}_{\ell',m',1/2,s}C^{J,M}_{\ell',m'',1/2,s}  \\ \nonumber
    & 
    \langle 
    n,\mathbf{k},s
    |r_{\mu}|
    \ell',m'
    \rangle 
    \langle 
    \ell',m''
    r_{\mu}|
    n,\mathbf{k},s|
    \rangle  \delta(\hbar \omega-\epsilon_{n \mathbf{k} s} + \epsilon_{\ell-1/2} ),
\end{align}
where we have neglected the variation of $\epsilon_J$ with respect to $\epsilon_{n
\mathbf{k} s}$, because the core energies are located at several thousands of
eV below the Fermi energy compared to our EXAFS range of a few hundreds of eV at the highest,
and as such we will now write $\epsilon_{\ell-1/2}=\epsilon_C$. Now, we can use
the orthogonality property of the Clebsch-Gordan coefficients:
\begin{align}
\label{closclebgor}
    &\sum_{J,M}C^{J,M}_{\ell',m',1/2,s}C^{J,M}_{\ell',m'',1/2,s}=\delta_{m',m''} \\
    \nonumber &
    \sum_{m_1,m_2}C^{J,,M}_{\ell_1,m_1,\ell_2,m_2}C^{J',M'}_{\ell_1,m_1,\ell_2,m_2}=\delta_{J,J'}\delta_{M,M'},
\end{align}
and write:
\begin{align}
   \sum_J \sigma_{J}^{\mu}(\omega) &= 4\pi \alpha 
    \sum_{m',n,\mathbf{k},s} (\epsilon_{n \mathbf{k} s} - 
    \epsilon_C) 
    \langle n,\mathbf{k},s |r_{\mu}| \ell',m' \rangle 
    \langle \ell',m' |r_{\mu}| n,\mathbf{k},s \rangle  
    \delta(\hbar \omega-\epsilon_{n \mathbf{k} s} + \epsilon_{C} ) \\ 
   \nonumber &=4\pi \alpha \hbar\omega \sum_{m',n,\mathbf{k},s} 
    \langle n,\mathbf{k},s |r_{\mu}| \ell',m' \rangle 
    \langle \ell',m' |r_{\mu}| n,\mathbf{k},s
    \rangle 
    \delta(\hbar \omega-\epsilon_{n \mathbf{k} s} + \epsilon_{C}).
\end{align}
Now, we can rewrite the Kohn-Sham eigenfunctions using the \textsc{PAW} method. For our purposes, we can remain at the partial wave contribution. We therefore have
\begin{equation}
   \sum_J \sigma_{J}^{\mu}(\omega) = 4\pi \alpha \hbar\omega \sum_{\substack{m',n,\mathbf{k},s \\ 
    p_1,\ell_1,p_2,\ell_2,m_1,m_2}} 
    \langle p_1,\ell_1,m_1 |r_{\mu}| \ell',m' \rangle 
    \langle \ell',m' |r_{\mu}| p_2,\ell_2,m_2\rangle
    P^*{}^{n,\mathbf{k},s}_{p_1,\ell_1,m_1}
       P^{n,\mathbf{k},s}_{p_2,\ell_2,m_2}
    \delta(\hbar \omega-\epsilon_{n \mathbf{k} s} + \epsilon_{C} ).
\end{equation}
Writing $r_{\mu}=r\sqrt{\frac{4
\pi}{3}}Y^{\mu}_1$, one can show that:
\begin{equation}
    \langle p_1,\ell_1,m_1 |r_{\mu}| \ell',m'
    \rangle =-\left( p_1,\ell_1|r|\ell'\right)C^{\ell',0}_{\ell_1,0,1,0}C^{\ell_1,m_1}_{1,\mu,\ell',m'},
\end{equation}
where the $\left( p_1,\ell_1|r|\ell'\right)$ are the radial integrations introduced in Eq. \ref{integWigEck}. 
This leads to:
\begin{align}
   \sum_J \sigma_{J}^{\mu}(\omega) &= 4\pi \alpha \hbar\omega \sum_{\substack{m',n,\mathbf{k},s \\ 
    p_1,p_2,\ell_1,\ell_2,m_1,m_2}} 
    \left( p_1,\ell_1 |r| \ell' \right) 
    \left( \ell' |r| p_2,\ell_2 \right) 
    C^{\ell',0}_{\ell_1,0,1,0}C^{\ell',0}_{\ell_2,0,1,0}
    C^{\ell_1,m_1}_{1,\mu,\ell',m'} C^{\ell_2,m_2}_{1,\mu,\ell',m'} \\ \nonumber 
    &\times P^*{}^{n,\mathbf{k},s}_{p_1,\ell_1,m_1}P^{n,\mathbf{k},s}_{p_2,\ell_2,m_2} 
    \delta(\hbar \omega-\epsilon_{n \mathbf{k} s} + \epsilon_{C} ).
\end{align}
We will now sum over all the polarization directions $\mu$ and use the orthogonality
relations from equation (\ref{closclebgor}):
\begin{align}
     \nonumber\sigma(\omega) = \sum_{\mu} \sum_J \sigma_{J}^{\mu}(\omega) &= 4\pi \alpha
     \hbar\omega \sum_{\substack{\mu,m',n,\mathbf{k},s \\ p_1,\ell_1,p_2,\ell_2,m_1,m_2}} 
     \left( 
     p_1,\ell_1 |r| \ell' \right) 
     \left( \ell' |r|p_2, \ell_2 \right) 
    C^{\ell',0}_{\ell_1,0,1,0}C^{\ell',0}_{\ell_2,0,1,0}
    C^{\ell_1,m_1}_{1,\mu,\ell',m'} C^{\ell_2,m_2}_{1,\mu,\ell',m'} \\ 
     \nonumber &\times 
     P^*{}^{n,\mathbf{k},s}_{p_1,\ell_1,m_1}
     P^{n,\mathbf{k},s}_{p_2,\ell_2,m_2} \delta(\hbar 
     \omega-\epsilon_{n \mathbf{k} s} + \epsilon_{C} ) \\
     \nonumber &= 4\pi \alpha \hbar\omega \sum_{\substack{n,\mathbf{k},s \\ 
     p_1,p_2,\ell_1,m_1}} 
     \left( 
     p_1,\ell_1 |r| \ell' \right) 
     \left( \ell' |r|p_2, \ell_2 \right)(C^{\ell_1,0}_{\ell',0,1,0})^2 \\
      &\times
     P^*{}^{n,\mathbf{k},s}_{p_1,\ell_1,m_1}
     P^{n,\mathbf{k},s}_{p_2,\ell_1,m_1}
     \delta(\hbar \omega-\epsilon_{n \mathbf{k} s} + \epsilon_{C} ).
\end{align}

Splitting the two allowed dipole transitions $\ell_1=\ell'\pm 1$, and neglecting the overlap between different projectors, we can rewrite this as:
\begin{align}
   \sigma(\omega)& \approx  4\pi \alpha \hbar\omega 
    \left[\left|(\ell'+1|r|\ell')C^{\ell',0}_{\ell'+1,0,1,0}\right|^2
    \rho_{\ell'+1}(\omega+\epsilon_C/\hbar)
    +\left|(\ell'-1|r|\ell')C^{\ell'-1,0}_{\ell',0,1,0}\right|^2\rho_{\ell'-1}(\omega+\epsilon_C/\hbar)\right] \\
   \nonumber &=4\pi \alpha \hbar (\omega'-\epsilon_C/\hbar)\left[\frac{\ell'+1}{2\ell'+3}
    \left|(\ell'+1|r|l')
    \right|^2\rho_{\ell'+1}(\omega')+\frac{\ell'}{2\ell'-1}\left|(\ell'-1|r|\ell')\right|^2\rho_{\ell'-1}(\omega')\right] \\
   \nonumber &=A_{\ell'}(\omega')\rho_{\ell'+1}(\omega')+B_{\ell'}(\omega')\rho_{\ell'-1}(\omega'),
\end{align}
where we introduced the $\ell$ partial densities of states $\rho_\ell$ and the
shifted frequencies $\omega'=\omega+\epsilon_C/\hbar$. Therefore, the
normalized edge $\sigma(\omega)$ can be written as a weighted sum of the
partial densities of states corresponding to the dipole allowed $\ell$ values.

\end{document}